\documentclass[10pt]{iopart}
\usepackage{graphicx}
\usepackage{dcolumn}
\usepackage{bm}

\usepackage{graphicx}
\usepackage{dcolumn}
\usepackage{bm}

\usepackage{amssymb}

\expandafter\let\csname equation*\endcsname\relax
\expandafter\let\csname endequation*\endcsname\relax
\usepackage{amsmath}

\usepackage{cancel}
\usepackage{braket}

\usepackage{float} 

\begin{document}

\title[The cell-centered Finite-Volume self-consistent approach for heterostructures:]{The cell-centered Finite-Volume self-consistent approach for heterostructures: 1D electron gas at the Si-SiO$_2$ interface}
\author{Vahid Mosallanejad \textsuperscript{1,*}, Haiou Li \textsuperscript{2}, Gong Cao \textsuperscript{2}, Kuei-Lin Chiu \textsuperscript{3,*}, Wenjie Dou \textsuperscript{1} and Guo-ping Guo \textsuperscript{2}.}
\address{$^1$ School of Science, Westlake University, Hangzhou, Zhejiang 310024, China $\&$ Institute of Natural Sciences, Westlake Institute for Advanced Study, Hangzhou, Zhejiang 310024, China}
\address{$^2$ CAS Key Laboratory of Quantum Information, and Synergetic Innovation Center of Quantum Information and Quantum Physics, University of Science and Technology of China,Chinese Academy of Sciences, Hefei 230026, China}
\address{$^3$ Department of Physics, National Sun Yat-Sen University, Kaohsiung 80424, Taiwan}
\ead{vahid@ustc.edu.cn and eins0728@gmail.com}
\vspace{10pt}
\begin{indented}
\item[] Jun 2023
\end{indented}

\date{\today}

\begin{abstract}
Achieving self-consistent convergence with the conventional effective-mass approach at ultra-low temperatures (below $4.2~K$) is a challenging task, which mostly lies in the discontinuities in material properties (e.g., effective-mass, electron affinity, dielectric constant). In this article, we develop a novel self-consistent approach based on cell-centered Finite-Volume discretization of the Sturm-Liouville form of the effective-mass Schr{\"o}dinger equation and generalized Poisson's equation (FV-SP). We apply this approach to simulate the one-dimensional electron gas (1DEG) formed at the Si-SiO$_2$ interface via a top gate. We find excellent self-consistent convergence from high to extremely low (as low as $50~mK$) temperatures. We further examine the solidity of FV-SP method by changing external variables such as the electrochemical potential and the accumulative top gate voltage. Our approach allows for counting electron-electron interactions. Our results demonstrate that FV-SP approach is a powerful tool to solve effective-mass Hamiltonians.
\end{abstract}


\vspace{2pc}
\maketitle
\ioptwocol
\label{sec:sample1}

\section{Introduction}
\label{sec1}
A comprehensive understanding of the electronic and optical properties of semiconductor heterostructures requires an accurate and efficient solution of the time-independent effective-mass Schr{\"o}dinger equation coupled with the Poisson's equation~\cite{ando1982electronic,duke1967optical, bloss1989effects,datta2005quantum}. However, solving such coupled equations is a difficult task: the time-independent effective-mass Schr{\"o}dinger equation is an eigenvalue problem, whereas Poisson's equation is an elliptic partial differential equation (PDE), such that analytical or simultaneous numerical solution is not available. The numerical self-consistent approach can be used to solve the coupled equations, which accurately predicts the electrostatic potential profile (\textit{band bending}) arising from various sources (such as ionized dopants, surface charges, and external gates)~\cite{yoshida1986classical,woods2018effective,davies1998physics}.
Such self-consistent solutions offer information on spatial-dependent observables, such as wavefunctions and electron density~\cite{degtyarev2017features}, which can be used to estimate the size of a quantum well~\cite{wang2004three,bell2010crossover}.

The underlying mathematics of the self-consistent Schr{\"o}dinger-Poisson (SP) field approach in semiconductor heterostructures is almost identical to the atomistic Hartree self-consistent (HSC) field theory, except for two major differences. Firstly, we have subbands in semiconductor heterostructures instead of orbitals in atomic HSC theory. Secondly, the number of electrons is not fixed in semiconductor heterostructures; Instead, the electrochemical potential (equivalent to Fermi-level) determines the occupation of subbands~\cite{stern1970iteration}. 
Due to the complexity of the calculations, the self-consistent SP approaches are largely carried out only in one dimension. Most studies focused on the two-dimensional electron gas (2DEG), where the effective dimension is the orientation perpendicular to the semiconductor heterojunctions~\cite{stern1972self,sarma1982electronic,lo1999modeling}. 
Finite-Difference Method (FDM) with uniform mesh (i.e., real-space basis set) has been taken as the primary numerical discretization method to solve the self-consistent coupled Schr{\"o}dinger-Poisson equations. FDM base self-consistent SP approaches (with uniform mesh) may suffer from convergence problems, especially for 2D problems due to the increasing number of basis sets~\cite {duarte2010convergence}. A nonuniform mesh can reduce the cost of 2D problems by reducing the number of the basis set~\cite{tan1990self,ando2002numerically}. In the late 80s and early 90s, Finite-Element Method (FEM) was introduced to semiconductor modeling~\cite{nakamura1989finite,wu1993self}.
Neither standard FEM nor FDM guarantee global and local conservations~\cite{mazumder2015numerical}. 

An alternative discretization approach is the Finite-Volume method~\cite{mazumder2015numerical}.
Besides ensuring conservation properties, the Finite-Volume method has a few unique advantages. These include (1) high stability, (2) systematic incorporation of material properties, and (3) self-validation.
In the context of device modeling Finite-Volume software is employed in two works to solve the Poisson equation~\cite{armagnat2019self, berrada2020nano}. In addition, the VSP software is, probably, the only implementation platform where the Voronoi-based Finite-Volume method is used for solving both Schr{\"o}dinger and Poisson equations~\cite{baumgartner2013vsp}. However, the implementation procedure is not discussed in detail. To the best of our knowledge, the cell-centered Finite-Volume method has not been applied to the self-consistent SP approach yet.   
The main focus of this study is to introduce the cell-centered Finite-Volume method for solving Schr{\"o}dinger-Poisson systems in semiconductor heterojunctions with particular attention to conservation laws and the implementation of a nonuniform mesh. 
We apply our method to 1DEG formed in the Si-SiO$_2$ heterostructure. Such a system is promising for the realization of scalable and high-fidelity spin qubits in low/ultra-low temperatures. We find good numerical results at temperatures as low as $50~mK$ insensitive to external variables such as top gate voltages. We can also include electron-electron interactions. Our approach suggests a complete treatment to extract a realistic confinement potential from the material and geometrical properties of the system with minimum assumptions.  

The paper is organized as follows. In subsection \ref{subsec21}, the 1DEG formed at the Si-SiO$_2$ interface is introduced. The subsections \ref{subsec22} and \ref{subsec23} explain how the cell-centered Finite-Volume method can be implemented to solve Schr{\"o}dinger and Poisson's equations, respectively. The Scaling of Schr{\"o}dinger and Poisson's equations are discussed in subsection \ref{subsec24}. The possibility of including many-body interactions into the problem is discussed in subsection \ref{subsec25}, while, the low-cost Finite-Volume Thomas-Fermi (FV-TF) approach is explained in subsection \ref{subsec26}. The cell-centered Finite-Volume predictor-corrector method, which accelerates the self-consistent field convergence, and its implementation are described in subsections \ref{subsec27} and \ref{subsec28}. Device and mesh geometries, solution convergence properties, and characteristics of 1DEG calculated by FV-TF and FV-SP will be presented in detail in Sec. \ref{sec2}. Benefits of the newly proposed approach and possible applications of this method will be summarized in Sec. \ref{sec3}.

\section{Methods}
\label{sec2} %
In this section, we introduce FV-SP approach. Note that our approach is general, but we restrict ourselves to the 1DEG (a 2D problem) described below. 

\subsection{Theory of one-dimensional electron gas}
\label{subsec21}
\begin{figure}[h]
	\begin{center}
		\includegraphics[width=8cm]{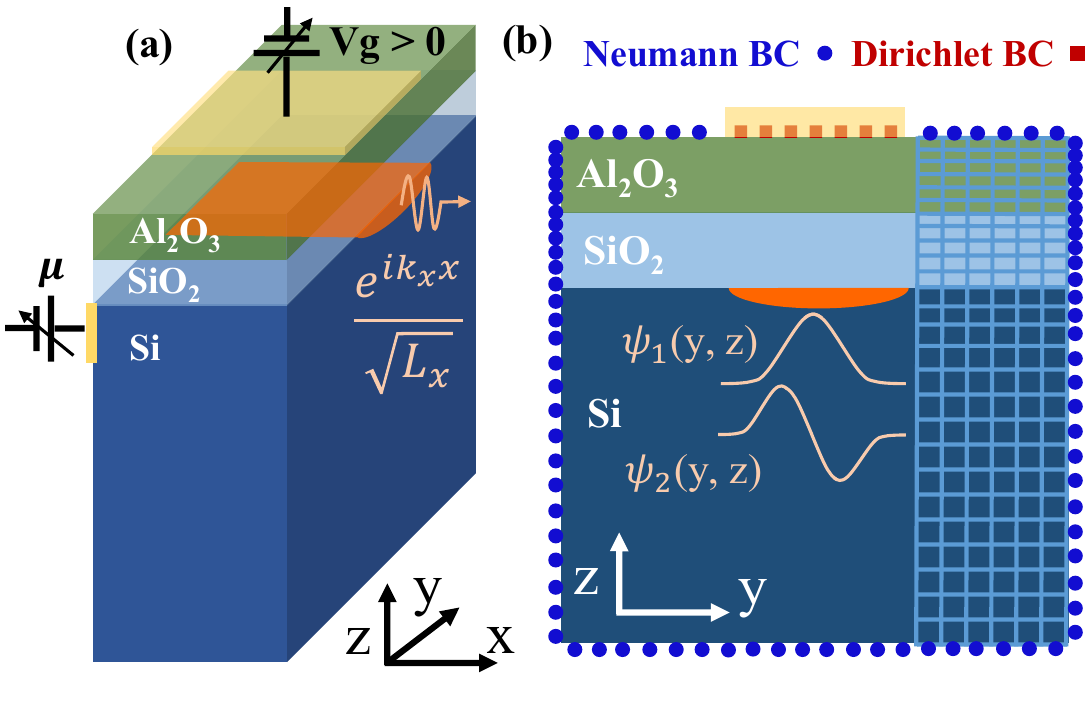}
	\end{center}
	\caption{\label{fig1} (a) A schematic of three-dimensional MOS quantum wire. One-dimensional electron gas is shown with orange color. (b) Using symmetry, the three-dimension geometry reduces to a two-dimension (2D) problem. A fixed Dirichlet Boundary Condition (BC) has been applied for the metal-insulator interface, red color cubic dots, and Neumann BC imposes for the rest, blue color circular dots.}
\end{figure}

We study the 1DEG formed in the three layers stack of Si-SiO$_2$-Al$_2$O$_3$ from bottom to top (shown in Fig. \ref{fig1}). We use such a model to mimic the experimental setup in Refs.~\cite{angus2007gate,brauns2018palladium}. Here, we do not consider extra transition layers. Assuming the 3D structure is uniformly periodic in the x-axis, and the 2D confinement exists on the yz-plane, the total wavefunction is given by
\begin{eqnarray}
\label{eq1}
\Psi (x,y,z)=\sum_{i}\frac{e^{ik_xx}}{\sqrt{L_x}}\psi_i(y,z),
\end{eqnarray}
where $k_x$ and $L_x$ are the subband wave vector and device length in the direction perpendicular to the 2D confinement, respectively~\cite{datta2005quantum}. 

In Eq. \eqref{eq1}, $\psi_i(y,z)$ is the subband wave function (equivalently known as envelope function), and $E_i$ is the subband energy, which together is determined by two-dimensional Sturm-Liouville form of one-band effective-mass Schr{\"o}dinger equation 
\begin{eqnarray}
\label{eq2}
\begin{aligned}
&\big[
\frac{\partial }{\partial y} 
\big(-\gamma_y(y,z)\frac{\partial }{\partial y}\big)+ 
\frac{\partial }{\partial z}
\big(-\gamma_z(y,z)\frac{\partial }{\partial z}\big)+U(y,z)
\big]\\
&\psi_i(y,z)=E_i~\psi_i(y,z).
\end{aligned}
\end{eqnarray}
In the above equation, $\gamma_{y,z}(y,z)$ is given by $\gamma_{y,z}(y,z)=\hbar^2/(2m_{y,z}^*(y,z))$~\cite{young1989position,levy1995position}. Here, $m^*_{y,z}$ refers to directional effective-mass, which is space-dependent. The Sturm-Liouville form of Schr{\"o}dinger equation is a Hermitian eigenproblem. Such a form allows for the effective-mass to vary over the space, which is the case for semiconductor heterojunctions. 
This form of the effective-mass Schr{\"o}dinger equation also preserves the continuity of probability current across junctions~\cite{burt1992justification}. $U(y,z)$ is the Hartree potential energy. In the simplest form, $U$ is the combination of electrostatic potential energy $-q\phi$, and band alignment discontinuity $\chi$ (electron affinity profile):
\begin{equation}
\label{eq3}
U(y,z)=-q\phi(y,z)+\chi(y,z),
\end{equation}
where $\phi$ and $q$ are the electrostatic potential and unit of charge, respectively. Note that, the above simple Hartree potential energy is related to the \textit{conduction band edge},  $E_c$, by $U=qE_c$. By definition, electron affinity $\chi$ is the amount of energy needed to push an electron from the bottom of the conduction band to the vacuum. In practice, $\chi$ is a known step-function (multi-steps across multi-layers and uniform along the other direction) available from experimental measurements or first-principle calculations~\cite{bersch2008band,ribeiro2009accurate}. In a more complex format, the exchange-correlation energy $U_{xc}$ accounting for many-body interactions can be added to the Hartree potential energy,
\begin{equation}
\label{eq4}
U(y,z)=-q\phi(y,z)+\chi(y,z)+U_{xc}(y,z).
\end{equation} 
$U_{xc}(y,z)$ will be discussed later. Here, we do not add the image charge potential to the Hartree potential $U$ explicitly~ \cite{stern1978image}. Such an effect has been also taken into account in the space-dependent dielectric constant within the generalized Poisson's equation. 
The generalized form of Poisson's equation is the correct equation which properly accounts for the static coulomb interactions in complex physical systems
\begin{eqnarray}
\begin{aligned}
\label{eq5}
&\big[
\frac{\partial }{\partial y} 
\big(-\epsilon_{ry}(y,z)\frac{\partial }{\partial y}\big)+ 
\frac{\partial }{\partial z} 
\big(-\epsilon_{rz}(y,z)\frac{\partial }{\partial z}\big)
\big] \phi(y,z) 
\\&
=\frac{-q}{\epsilon_0}n(y,z).
\end{aligned}
\end{eqnarray}
The derivative with respect to $x$ vanishes due to the uniformity of electrostatic potential in the x-axis. We have taken into account the anisotropy and space dependency for the directional relative static dielectric constants $\epsilon_{ry,~rz}$. This form of Poisson's equation preserves the continuity of the electric displacement across Si-SiO$_2$ and SiO$_2$-Al$_2$O$_3$ interfaces. In order to reduce the complexity, we did not include any doping (or partial ionization), surface charge, and dipole as sources of charge in Poisson's equation~\cite{snider1990electron,mizsei2002fermi,kamata2017design}. 
The electron density is evaluated using the wavefunctions from the Schr{\"o}dinger equation. The quantum electron density defines as $n(y,z)= L_x^{-1}\sum_{i,k_x}
|\psi_i(y,z)|^{2}$ $ f_0(E_i+\hbar^2 k_x^2/2m_x-\mu)$, where $f_0$ is the Fermi-Dirac function. In the presence of 2D confinement, the quantum electron density per valley and per spin can be expressed as:
\begin{equation}
\begin{split}
\label{eq6}
n(y,z)
&=\frac{\sqrt{m_x(y,z)}}{\sqrt{2}\pi\hbar}
\sum_i|\psi_i(y,z)|^2
\int_{E_i}^{\infty} 
\frac{(\epsilon-E_i)^{-\frac{1}{2}} d\epsilon}{1+exp(\frac{\epsilon-\mu}{k_BT})}\\
&=\frac{\sqrt{m_x(y,z)k_BT}}{\sqrt{2\pi}\hbar}
\sum_i|\psi_i(y,z)|^2 \mathcal{F}_{-\frac{1}{2}}
\Big(\frac{\mu-E_i}{k_BT} \Big),
\end{split}
\end{equation}
where $m_x(y,z)$ is the electron's effective mass on the x-axis, and $\mu$ refers to the electrochemical potential which is a fixed value here \cite{pacelli1997self,heinz2004simulation}. In fact, we have assumed the 2D heterostructure is connected to an external source of particles. Note that, there are theoretical works studying isolated (modulated doped) heterostructures where $\mu$ is taken as a variable which is determined by the number of charges~\cite{ram2004schrodinger}. The factor $\mathcal{F}_{-{1}/{2}}$ stands for the complete Fermi-Dirac integral of the order $-1/2$. Indeed, $\sqrt{m_x} / \pi\hbar\sqrt{2 (\epsilon-E_i)}$ is the standard density of state for a quantum wire. The exact form of the complete Fermi-Dirac integral of order $j$ is given by
\begin{eqnarray}
\begin{aligned}
\label{eq7}
\mathcal{F}_j(\eta)=\frac{1}{\Gamma^f(j+1)}\int_{0}^{\infty}
\frac{\varepsilon^j d\varepsilon}{1+exp(\varepsilon-\eta)},
\end{aligned}
\end{eqnarray}
where $\Gamma^f$ is the gamma function~\cite{lether2000analytical}.  
Note that, summing over all wave vectors $k_x$ give rise to the first expression in Eq. (\ref{eq6}) while changing the integration variable results in the second. We do not include any phenomenological terms in our electronic density. The electron density given in Eq. (\ref{eq6}) has three contributions as (I) an effective one-dimensional density, $N_{C_{1D}}(y,z)\equiv \sqrt{m_x(y,z)} (2\pi)^{-1/2}\hbar^{-1}{(k_BT)}^{1/2}$ (with the physical unit of length$^{-1}$), (II) the probability density $|\psi_i(y,z)|^2$ (with the physical unit of length$^{-2}$), and (III) the factor $\mathcal{F}_{-{1}/{2}}(\frac{\mu-E_i}{k_BT})$, which can be referenced as the subband's occupancy factor. In fact, values of $\mathcal{F}_{-1/2}(E_i)$ tell us which subbands are filled and which are empty.

$\mathcal{F}_{j}(\eta)$ has the derivative property $\partial \mathcal{F}_{j}(\eta)/ \partial \eta =j \mathcal{F}_{j-1}(\eta)$ \cite{kim2008notes}. In addition, the factor $\mathcal{F}_{{1}/{2}}(x)$ can be approximated by a function \cite{bednarczyk1978approximation,gao2013quantum}. $\mathcal{F}_{{1}/{2}}(x) $ approaches to $ exp(x)$ at $x\rightarrow -\infty$, and $4 x^{3/2}/3\sqrt{\pi}$ at $x\rightarrow +\infty$ such that $\mathcal{F}_{{-1}/{2}}(x) \propto exp(x)$ at $x\rightarrow -\infty$, and $\mathcal{F}_{{-1}/{2}}(x) \propto x^{1/2}$ at $x\rightarrow +\infty$. Consequently, at the limit of zero temperature, we can expect vanishing contributions to the electron density from $E_i>\mu$ while the contributions of $E_i<\mu$ became temperature independent. In such a regime, it is essential to obtain an accurate numerical evaluation of the Fermi-Dirac integrals rather than analytical approximations. Roughly speaking, the i-th component of electron density is nonzero if the subband energy $E_i$ is below the $\mu$. Moreover, the number of states below $\mu$ and their energy distances to $\mu$ can not be predetermined. Hence, an analytical estimation of the electron density can not be provided when $T\rightarrow 0$. A combination of Eq. (\ref{eq2}), Eq. (\ref{eq5}), and Eq. (\ref{eq6}) is the coupled Schr{\"o}dinger Poisson equations we intended to solve. Below, we introduce the discretization scheme to solve these equations numerically. 

\subsection{Cell-centered discretization scheme in 2D}
\label{subsec22}
\begin{figure}[]
	\begin{center}
		\includegraphics[width=8.5cm]{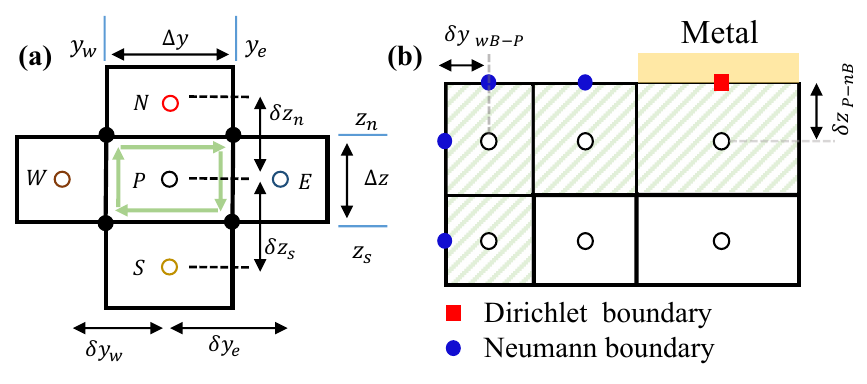}
	\end{center}
	\caption{(a) Arrangement of structured cell-centered Finite-Volume grids. The central cell ($P$) has four neighbor cells. Central coordinates are labeled as $P$, \textit{N}, \textit{S}, \textit{E}, and \textit{W} and are shown with hollow circles. Nodal coordinates are shown with the filled circles. (b) A schematic of boundary cells and boundary points on the northwest corner of a rectangular domain. Patterned cells are boundary cells, while non-patterned cells are regular cells. Blue-color/filled circular markers are Neumann boundary points, whereas the red-color/filled cubic marker is a Dirichlet boundary point.}
 \label{fig2} 
\end{figure}

Cell-centered Finite-Volume scheme has been chosen among a few arrangements of Finite-Volume methods \cite{blazek2015computational}. The first step in Finite-Volume discretization is to integrate over a \textit{control volume} (CV). The concept \textit{control volume} commonly refers to a Finite-Volume cell, the central cell in Fig. \ref{fig2}(a). In the structured cell-centered Finite-Volume, each CV has four neighbor CVs on the Northern, Southern, Eastern, and Western sides. The cell's center is labeled as \textit{P} in the Fig. \ref{fig2}(a), and neighbors are labeled as \textit{N}, \textit{S}, \textit{E}, and \textit{W}.
Both the Schr{\"o}dinger equation in Eq. \eqref{eq2} and Poisson's equation in Eq. \eqref{eq5} have a Laplacian operator as follows: 
\begin{eqnarray}
\label{eq8}
&\frac{\partial }{\partial y}
\Big(\Gamma_y(y,z)\frac{\partial }{\partial y}\Big)+
\frac{\partial }{\partial z}
\Big(\Gamma_z(y,z)\frac{\partial }{\partial z}\Big). 
\end{eqnarray}
The $\Gamma$ is commonly called the \textit{diffusion coefficient}~\cite{chen2010new}. In Poisson's equation, $\Gamma$ represents the dielectric constant (relative permittivity). Taking Poisson's equation, we start with approximating the integration of the Laplacian operator that acts on the electrostatic potential, $\phi$ as 
\begin{eqnarray}
\label{eq9}
\begin{aligned}
&\int_{z_s}^{z_n}\int_{y_w}^{y_e}
\bigl[
\frac{\partial}{\partial y}
\Big(\Gamma_y \frac{\partial \phi}{\partial y}\Big)
+\frac{\partial}{\partial z}
\Big(\Gamma_z \frac{\partial \phi}{\partial z}\Big)
\bigr]dydz
=\int_{z_s}^{z_n} \bigl[\Gamma_y
\\&
 \frac{\partial \phi}{\partial y}\Bigr\rvert_{y_e}-
          \Gamma_y \frac{\partial \phi}{\partial y}\Bigr\rvert_{y_w} \bigr]dz
+\int_{y_w}^{y_e}\bigl[\Gamma_z \frac{\partial \phi}{\partial z}\Bigr\rvert_{z_n}-
       \Gamma_z \frac{\partial \phi}{\partial z}\Bigr\rvert_{z_s} \bigr]dy
\\&
\approx 
\big[\Gamma_y \frac{\partial \phi}{\partial y}\Bigr\rvert_{y_e}-
\Gamma_y
\frac{\partial \phi}{\partial y}\Bigr\rvert_{y_w}\big]\Delta z+ 
\\&
\big[\Gamma_z \frac{\partial \phi}{\partial z}\Bigr\rvert_{z_n}-\Gamma_z 
\frac{\partial \phi}{\partial z}\Bigr\rvert_{z_s}\big]  \Delta y
\\&
\approx \bigl[ 
\Gamma_{y_e} \frac{ \phi_E-\phi_P}{\delta y_e}-
\Gamma_{y_w} \frac{ \phi_P-\phi_W}{\delta y_w} 
\bigr] \Delta z+ \\
&
~~\bigl[ 
\Gamma_{z_n} \frac{ \phi_N-\phi_P}{\delta z_n}-
\Gamma_{z_s} \frac{ \phi_P-\phi_S}{\delta z_s} 
\bigr] \Delta y.
\end{aligned}
\end{eqnarray}
In the above approximation, analytical integration has been used to derive the line-integral terms. In the next step, we have assumed that new integrands (flux) do not change along integration paths. Equivalently, the average flux is assumed to be identical to the value of the flux computed at the center of the face~\cite{mazumder2015numerical}. It is important to note that, the partial use of analytical integration will improve the achievable accuracy of the Finite-Volume method. The physical interpretation of Eq. \eqref{eq9} is given by the Divergence Theorem as: the surface integral of the divergence of the electric displacement field over a CV has been replaced with the line integral of normal components of the electric displacement along close boundaries of that CV. In the last line of the above equation, we have used the central difference approximation to the derivatives on interfaces. 
For the right-hand side of Poisson's equation, we approximate the integration of the electron density $n(y,z)$ [the source term] over a CV  with a piecewise constant: 
\begin{equation}
\label{eq10}
\int_{z_s}^{z_n}\int_{y_w}^{y_e} n(y,z)dydz\approx n_P\Delta y \Delta z.
\end{equation}
This rather crude approximation allows us to keep the conservative nature of cell-centered Finite-Volume. The above approximation is interpreted with the following assumption: the average value of electron density over a cell is approximated by the local value of electron density at the cell’s center. Therefore, different from FDM or FEM, the Finite-Volume method has the closest connection to experimental measurements: the parameter of interest [e.g., the local density of states (LDOS)] is measured locally as an average value in a specific coordinate. The Finite-Volume method is thus considered to be a physical method for solving PDEs rather than a purely mathematical method.     
Substituting the approximations given in Eqs. \eqref{eq9} and \eqref{eq10} into Eq. \eqref{eq5} and dividing both sides by $\Delta y \Delta z$, we arrive at a discrete Poisson's equation:
\begin{eqnarray}
\label{eq11}
\begin{aligned}
&-a_W \phi_W +[-a_S \phi_S+a_P \phi_P-a_N \phi_N] -a_E \phi_E
\\&
= -\frac{q}{\epsilon_0}n_P, ~~~~ a_P=a_W+a_S+a_N+a_E,
\end{aligned}
\end{eqnarray}
where we have defined coefficients, $a_W=\Gamma_{yw}/ \delta y_w \Delta y$, $a_E=\Gamma_{ye}/ \delta y_e \Delta y$, $a_S=\Gamma_{zs}/ \delta z_s \Delta z$, and $a_N=\Gamma_{zn}/ \delta z_n \Delta z$. We refer to these coefficients as \textit{a-coefficient}s. The correct method to incorporate the parameter $\Gamma$s (e.g., $\Gamma_{yw}$) at the interfaces into a-coefficients is extremely vital to maintain the conservation laws on the local level. To achieve the conversation for certain quantities, flux continuities at four interfaces are enforced by taking a \textit{harmonic mean} approximation for the value of $\Gamma$s at the interfaces. To be more specific, we approximate $\Gamma_{ye}$ as $(\Gamma_E\Gamma_P)/(\beta\Gamma_E+(1-\beta)\Gamma_P)$ to keep the continuity of the flux at the CV's Eastern interface. Here $\beta=\delta y_{e^-}/\delta y_{e}$, $(1-\beta)= \delta y_{e^+}/\delta y_{e}$ and $\delta y_{e^-}$ ($\delta y_{e^+}$) is the distance between the $P$ ($E$) to the eastern interface. A similar approximation is used for the other $\Gamma$s. Derivation of the harmonic mean approximation is explained in \ref{AppA}.
The kinetic energy operator in the Schr{\"o}dinger equation has a similar derivative operator as in Poisson's equation. With similar procedures, we can derive the following discrete equation for the effective-mass Schr{\"o}dinger equation:
\begin{eqnarray}
\label{eq12}
\begin{aligned}
&-a'_W {\psi_i}_W -a'_S {\psi_i}_S+a'_P {\psi_i}_P-a'_N {\psi_i}_N -a'_E {\psi_i}_E
\\&
=E_i{\psi_i}_P,~~~~a'_P=a'_W+a'_S+a'_N+a'_E+U_P,
\end{aligned}
\end{eqnarray}
where 
$a'_W=\gamma_{yw}/ \delta y_w \Delta y$, $a'_E=\gamma_{ye}/ \delta y_e \Delta y$, $a'_S=\gamma_{zs}/ \delta z_s \Delta z$, and $a'_N=\gamma_{zn}/ \delta z_n \Delta z$. 
The integrals of the source, and potential energy terms,  ($\int_{z_s}^{z_n}\int_{y_w}^{y_e}$ $E_i\psi(y,z)dydz, \int_{z_s}^{z_n}\int_{y_w}^{y_e}U(y,z)\psi(y,z)dydz$), are approximated with $E_i{\psi_i}_P\Delta y \Delta z$ and $U_P{\psi_i}_P\Delta y \Delta z$, respectively. We point out that terms $a'_{W, S, E, N}$ serve as \textit{hopping energies} in standard tight-binding Hamiltonian models, and the $a'_P$ serves as the \textit{on-site potential energies}. Different from the tight-binding Hamiltonian, these coefficients are not fixed values due to the nonuniform mesh. 
Discrete equations, Eqs. \eqref{eq11} and \eqref{eq12}, must run over all cells in the domain. For the discrete Poisson's equation, if the $n_P$ is known, the system of linear algebraic equations can be arranged in the following matrix form: $[A] \vec{\phi}~=\vec{n}+\vec{R_B}$. The pattern of a-coefficients matrix, $[A]$, depends on the so-called \textit{global ordering}, which is referred to the order of all cells in the computation domain. 
The vector $\vec{R_B}$ is referred to as a vector consisting of fixed values on (Dirichlet) boundary faces which is transferred to the right side of the equality. In the following subsection, we give a clear explanation of how boundary conditions are implemented within the framework of cell-centered Finite-Volume. Nevertheless, knowing the $\vec{n}$ (on-site electron density), one can calculate the profile of electrostatic potential with direct diagonalization $\vec{\phi}=[A]^{-1}(\vec{n}+\vec{R_B})$. Indeed, $\vec{n}$ is an unknown function of $\vec{\phi}$ which should be calculated self-consistently. Later (in subsections \ref{subsec26} and \ref{subsec27}), we will elaborate on how one can estimate ${n_P}$ in terms of ${\phi_P}$ in the framework of the Finite-Volume method.
In the case of the discrete Schr{\"o}dinger equation, the system of linear algebraic equations can be compacted in a matrix form as $[H]~\vec{{\psi_i}}~=E_i\vec{{\psi_i}}$. 

\subsection{Implementation of boundary conditions}
\label{subsec23}
We start with the implementation of boundary conditions for Poisson's equation. There are two types of boundary conditions for Eq. \eqref{eq11}. Dirichlet boundary condition must be implemented on cells that are connected to the top metal gate, where the quantity of electrostatic potential is a known value (see the red color/filled cubic marker in Fig. \ref{fig1}). Whereas, the Neumann zero flux boundary condition must be implemented elsewhere (depicted by blue-color/filled circular markers in Fig. \ref{fig1}). In what follows, we refer to them as \textit{boundary point}s. A few boundary CVs are highlighted with a light pattern on a schematic illustration in Fig. \ref{fig2}(b), where we show a few meshes on the northwest corner of a hypothetical rectangular domain. Regardless of the type of PDE, Finite-Volume cells can be divided into two categories: (I) \textit{regular CV}s, Cells that have all their four faces connected to other cells. (II) \textit{boundary CV}s, Cell that one or two of their faces are connected to the environment (vacuum or metal in our problem). For boundary CVs, the approximation of derivatives associated with line integrals along boundary interfaces must be modified based on the type of boundary condition, i.e., a-coefficient must be appropriately corrected for those CVs which belong to the category of boundary CVs.

The implementation of the Dirichlet boundary condition for the north-most cell whose northern face is attached to the metal is as follows. The first-order derivative operator (electric displacement) associated with the line integral along the north face is approximated differently via 
\begin{equation}
\label{eq13}
\Gamma_z \frac{\partial \phi}{\partial z}\Bigr\rvert_{z=z_{nB}}=
\Gamma_{z_{nB}} \frac{ \phi_{nB} -\phi_P}{\delta z_{P-nB}},
\end{equation}
where $\phi_{nB}=V_g$ (top gate voltage) is a known value. The $\delta z_{P-nB}$ refers to the distance between the center of the northernmost cell and the closest northern boundary point and $\phi_{nB}\equiv\phi(y_{P},z_{nB})$ and $\phi_P\equiv\phi(y_{P},z_{P})$. We note that $\Gamma_{z_{nB}}$ refers to the material property (relative dielectric permittivity in Poisson's equation, effective-mass in the Shr\"odinger's equation) on the closest northern boundary points and we do not need to calculate $\Gamma_{z_{nB}}$ with the harmonic mean (\ref{AppA}). 

The consequences of the new approximation [i.e., Eq. \eqref{eq13}] on Eq. \eqref{eq11} are threefold. Firstly, a partially non-zero vector, $\vec{R_{nB}}=\Gamma_{z_{nB}} \vec{\phi_{nB}} /\delta z_{P-nB}$ must be transferred to the right-hand side when solving the linear algebraic equation. 
In practice, each boundary CV of the domain should be investigated whether the Dirichlet boundary condition is applied to each of the four faces. In cases when non-vanishing boundary potentials are connected to the other respective faces of the domain, vectors of $\vec{R_{sB}}$, $\vec{R_{wB}}$ and $\vec{R_{eB}}$ must be constructed and the total vector $\vec{R_B}=\vec{R_{sB}}+\vec{R_{nB}}+\vec{R_{wB}}+\vec{R_{eB}}$ which contains all Dirichlet boundary information, must be transferred to the right-hand side of the system of linear algebraic equations. 
In our geometry, i.e., Fig. \ref{fig1}(b), the $\vec{R_B}$ reduces to only $R_{nB}$ because the boundary condition for all other interfaces is Neumann. Secondly, the appeared $a_N$ on Eq. \eqref{eq11} must vanish. Thirdly, the quantity $a_N$ on the diagonal elements, $a_P=(a_W+a_S+a_{NB}+a_E)$, must be replaced with $a_{NB}=\Gamma_{z_{nB}} /(\delta z_{P-nB}\Delta z)$. 
The implementation of the Neumann boundary condition is effortless. The zero-flux condition for the northern face of a north-most cell implies
\begin{equation}
\label{eq14}
\Gamma_z \frac{\partial \phi}{\partial z}\Bigr\rvert_{z=z_{nB}}=0.
\end{equation}
Looking at Eq. \eqref{eq9}, one can perceive that the consequence of the zero flux boundary condition on Eq. \eqref{eq11} leads to a vanishing $a_N$ both appeared in Eq. \eqref{eq11} and in the $a_P=(a_W+a_S+\cancel{a_{N}}+a_E)$. Similar vanishing action implements the Neumann boundary condition at other Western, Eastern, and Southern interfaces. 

For the Schr{\"o}dinger equation, there is only one boundary condition, i.e., zero Dirichlet boundary condition on all outermost surfaces, since all subband wavefunctions converge to zero in the vacuum. As a result, two modifications must take place during the construction of the Hamiltonian matrix (using a$^\prime$-coefficients) such that if the cell has one or two boundary faces, the corresponding a$^\prime$-coefficient in Eq. \eqref{eq12} and the corresponding a$^\prime$-coefficient in the $a^\prime_P$ must vanish. Therefore, there is nothing to be transferred to the right side in the process of correcting the Hamiltonian matrix, and hence the implementation of boundary condition for Schr{\"o}dinger equation only modifies the diagonal elements of the Hamiltonian matrix. For this reason, the matrix form of Schr{\"o}dinger equation can be given by $[H]~\vec{{\psi_i}}~=E_i\vec{{\psi_i}}$.
  
\subsection{Scaling} 
\label{subsec24} 
So far, we have explained how a-coefficients are calculated and being corrected (for different boundary conditions) for our original Schr{\"o}dinger and Poisson's equations. However,  It is more practical to calculate and correct a$^\prime$-coefficient and a-coefficients for the scaled Schr{\"o}dinger and Poisson's equations. To do so, we first define a unified length scale $L_{sc}$ and scale the coordinates as $y\rightarrow L_{sc}y$ and $z\rightarrow L_{sc}z$. In our calculation, we choose $L_{sc}=10^{-9}$ such that we can rewrite the  Schr{\"o}dinger equation as follows: 
\begin{eqnarray}
\begin{aligned}
\label{eq15}
&\bigl[
\frac{\partial }{\partial y} 
\Big(-\frac{1}{r^{\ast}_y(y,z)}\frac{\partial }{\partial y}\Big)+ 
\frac{\partial }{\partial z}
\Big(-\frac{1}{r^{\ast}_z(y,z)}\frac{\partial }{\partial z}\Big)+E_{sc}
\bigr]~~~~
\\
&\psi_i(y,z)=S_i~\psi_i(y,z).
\end{aligned}
\end{eqnarray}
In the above scaled equation, $r^{\ast}_{y,z}=m^{\ast}_{y,z}/m_0$ and $E_{sc}$ and $S_i$ denote the scaled conduction band edge and the scaled eigenvalues as:
\begin{eqnarray}
\begin{aligned}
\label{eq16}
&E_{sc}(y,z)=\frac{(-\phi(y,z)+\chi_e(y,z))}{V_{sc}},
~
S_{i}=\frac{e_i}{V_{sc}}.
\end{aligned}
\end{eqnarray}
Here, $V_{sc}=\hbar^2 /(2m_0qL^2_{sc})$, which has a physical unit of $eV.m^2$, and we define $\chi_e=\chi/q$ and $e_i=E_i/q$.
For the scaled Schr{\"o}dinger equation, scaled a$^\prime$-coefficients are: $a^\prime_W={(r^{\ast}_{yw} \delta y_w \Delta y)}^{-1}$, $a^\prime_E={(r^{\ast}_{ye} \delta y_e \Delta y)}^{-1}$, $a^\prime_S={(r^{\ast}_{zs} \delta z_s \Delta z)}^{-1}$, and $a^\prime_N={(r^{\ast}_{zn} \delta z_s \Delta z)}^{-1}$. 
We denote the matrix form of the scaled Schr{\"o}dinger equation as: $[H_{sc}]~\vec{{\psi_i}}~=S_i\vec{{\psi_i}}$. The benefit of the scaled Schr{\"o}dinger equation is two-fold. Firstly, geometry definition and meshing will be carried out in the larger unit (meter). Secondly, Hamiltonian matrix elements (scaled a$^\prime$-coefficient) are neither excessively small nor excessively large (e.g., $a^{\prime}_W=400~m^{-2}$  by taking $r^{\ast}_{yw}=0.25, \delta y_w=\Delta y=0.1~m$). Therefore, the calculation of eigenvalues and eigenfunctions will be less negatively influenced by numerical errors that arise due to excessively small a-coefficients in the Hamiltonian matrix. The scaled Poisson's equation reads as
\begin{eqnarray}
\begin{aligned}
\label{eq17}
&\bigl[
\frac{\partial }{\partial y} 
\Big(\epsilon_{ry}(y,z)\frac{\partial }{\partial y}\Big)+ 
\frac{\partial }{\partial z} 
\Big(\epsilon_{rz}(y,z)\frac{\partial }{\partial z}\Big)
\bigr]
\phi(y,z)=
\\&
\frac{L^2_{sc}q}{\epsilon_0}n(y,z).
\end{aligned}
\end{eqnarray}
By taking the same scale length, $L_{sc}=10^{-9}$, the scale factor for Poisson's equation is $L^2_{sc}q/\epsilon_0\approx$ $10^{-26}~V.m^3$. Note that, the order of two-dimensional electron density is roughly around  $10^{24}~m^{-3}$~\cite{davies1998physics}. Such that the order of the magnitude for the right-hand side of the above equation is $10^{-2}$, which is well in agreement with having a range of several tens of meV for the depth of quantum well [note that, $E_c(y,z)$ is proportional to $-\phi(y,z)$]. 

\subsection{Electron-electron interaction}   
\label{subsec25}
To include electron-electron interaction, one should add the exchange-correlation energy in eV unit, $V_{xc}=U_{xc}/q$, to the numerator of Eq. \eqref{eq16} as $E_{sc}(y,z)=(-\phi(y,z)+\chi_e(y,z)+V_{xc}(y,z)~)/V_{sc}$. The local exchange-correlation potential can be expressed as
\begin{eqnarray}
\begin{aligned}
\label{eq18}
&V_{xc}(y,z)=-\frac{q^3}{32 \pi^3 \hbar^2}
\big(\frac{9 \pi}{4 }\big)^{{1/3}}
\frac{m^\star(y,z)}{{\epsilon^\star(y,z)}^2} \times \\
&~~~~~~~~~~~~~~ 
\bigl[ I^\star(y,z)+0.0545~log 
\big(1+11.4I^\star(y,z)~\big)\bigr],\\
&
I^\star (y,z)\equiv \frac{1}{r^\star(y,z)}=\big({\frac{4\pi}{3}}\big)^{1/3}  a^\star(y,z){n(y,z)}^{1/3},\\
& a^\star(y,z) =\frac {4\pi \hbar^2 }{q^2}
  \frac{\epsilon^\star(y,z)}{m^\star(y,z)}, \quad
   \frac{1}{m^\star(y,z)}=\frac{1}{3}
   \Big( 
   \frac{1}{m_l}+\frac{2}{m_t}
   \Big),
\end{aligned}
\end{eqnarray}
where $r^\star$ is the average distance between charges and $a^\star$ is the effective Bohr radius and $m_t$ ($m_l$) refers to the transverse (longitudinal) effective-mass~\cite{ando2003self}. Note that, $V_{xc}$ commonly is expressed in the SI system in literature, whereas here it refers to the exchange-correlation potential in the eV unit. The above $V_{xc}$ intentionally has been expressed in terms of $I^\star$, i.e., the inverse of the average distance between charges. The seminal expression for $V_{xc}$ is given in terms of the inverse of electron density, such as the expression given in Ref. ~\cite{hedin1971explicit}. 
Therefore, the value of any parameters in terms of the inverse of electron density may exceed the realizable floating-point number of the computation machine. Slightly different exchange-correlation functional suggested by Hedin and Lundqvist can also be rearranged in the same fashion as
\begin{eqnarray}
\begin{aligned}
\label{eq19}
&V_{xc}(y,z)=-\frac{q^3}{16 \pi^3 \hbar^2}
\big(\frac{9 \pi}{4 }\big)^{{1/3}}
\frac{m^\star(y,z)}{{\epsilon^\star(y,z)}^2} \times \\
&~~~~~~~~~~~~~~~
\bigl[I^\star(y,z)+0.0368~log \left(1+21I^\star(y,z) \right)\bigr],\\
\end{aligned}
\end{eqnarray}
with the same definitions for $I^\star$ and effective Bohr radius $a^\star$~\cite{stern1984electron,gao2013quantum}. Even though the order of parameters is the same in Eq. \eqref{eq18} and Eq. \eqref{eq19}, it is evident that these two equations differ in the three constant values [e.g., the 11.4 in Eq. \eqref{eq18} vs the 21 in Eq. \eqref{eq19}]. In the next subsection, we will explain how the semi-classical Thomas-Fermi approximation can provide fairly reasonable initial guesses for both electrostatic potential and electron density required for the first cycle of the self-consistent solution.
\subsection{Thomas-Fermi approximation}  
\label{subsec26} 
The semi-classical Thomas-Fermi approximation refers to assuming a plane-wave form for the wavefunction. With this approximation, the electron density obtains by summing the 3D normalized plane waves that occupy the contentious energy spectrum starting from the bottom of the conduction band~\cite{datta2005quantum}. Excluding electron-electron interaction, the general Thomas-Fermi approximation for electron density per valley per spin is expressed as:
\begin{equation}
\label{eq20}
n(\vec{r})=
\frac{\sqrt{m_x^\star m_y^\star m_z^\star}}{{\pi^{3/2}}~~\hbar^3}
{(k_BT)}^{\frac{3}{2}} \mathcal{F}_{\frac{1}{2}}\Big(\frac{\mu_e+(\phi-\chi_e)}{V_T} \Big),
\end{equation}
where $(\vec{r})=(x,y,z)$, and $\phi(\vec{r})$ is the unknown electrostatic potential. Hereafter, we define $\mu_e=\mu/q$ which is a fixed electrochemical potential in eV unit~\cite{heinz2004simulation}. Note that, the bottom of the conduction band edge is: $Ec=-\phi+\chi_e$. The parameter $V_T=k_BT/q$ is called the thermal energy. The function $\mathcal{F}_{{1}/{2}}$ stands for the complete Fermi-Dirac integral of the order $1/2$. The term $N_C(\vec{r})= \sqrt{m_x^\star m_y^\star m_z^\star}~\pi^{-3/2}~\hbar^{-3}{(k_BT)}^{3/2}$, is the effective density of state for 
free electrons in the bulk of a semiconductor with anisotropic effective-masses. Note that, variables $\phi$, $\chi_e$ (and hence $Ec$) and $m_{x, y, z}^\star$ are space dependent only on the $(y,z)$ in our 2D problem. One can substitute the above approximation into Eq. \eqref{eq17}, to form a nonlinear scaled Poisson's equation. Solving this nonlinear PDE provides rough estimations for profiles of electrostatic potential and electron density which self-consistent SP cycle. 
As the first step of the Finite-Volume method, one must approximate the integration of the right- and left-hand sides of the PDE over a CV. The integral approximation of the nonlinear source term on the right side is given by
\begin{eqnarray}
\begin{aligned}
\label{eq21}
\int_{z_s}^{z_n}\int_{y_w}^{y_e} 
&N_C(y,z) \mathcal{F}_{\frac{1}{2}}
\Big(\frac{\mu_e+\phi(y,z)-\chi(y,z)}{V_T}
\Big)dydz 
\\&
\approx
N_{C_P}~\mathcal{F}_{\frac{1}{2}}
\Big(\frac{\mu_e+\phi_P-\chi_P}{V_T}\Big)
\Delta y \Delta z.
\end{aligned}
\end{eqnarray}
In the above relation, an average of the integrand (over a CV) is given by substituting the average values of the variables (the average density of state, $N_{C_P}$, the average electrostatic potential, $\phi_P$, and the average electron affinity, $\chi_P$) in the integrand multiplying by the cell volume. This assumption becomes better as the size of the CV decreases. We approximate the integration of the Laplacian operator, in the left-hand side of Eq. \eqref{eq17}, by the a-coefficient times to the cell volume. The discretized form is obtained by equating approximations of the right- and  the left-hand sides as the following
\begin{eqnarray}
\begin{aligned}
\label{eq22}
&-a_W \phi_W +[-a_S \phi_S+a_P \phi_P-a_N \phi_N] -a_E \phi_E=
\\&
-\frac{\nu_v\nu_sq L^2_{sc}}{\epsilon_0}N_{C_p}\mathcal{F}_{\frac{1}{2}}
\Big(\frac{\mu_e+\phi_P-\chi_P}{V_T}\Big),
\end{aligned}
\end{eqnarray}
where we have also added the valley, $\nu_v$, and spin, $\nu_s$, degeneracies. After running Eq. \eqref{eq22} for all CVs and the proper implementation of boundary conditions, the residual form of the system of nonlinear algebraic equations reads as:
\begin{equation}
\label{eq23}
\vec{d}~(\vec{\phi})=[A]~\vec{\phi}~+\frac{\nu_v\nu_sq L^2_{sc}}{\epsilon_0}\vec{N}_{C} \vec{\mathcal{F}}_{\frac{1}{2}}
(\vec{\phi},\vec{\chi},\mu_e,V_T)-\vec{R}_B=0.
\end{equation}
The matrix $[A ]$ and the vector $\vec{R}_B$ are the same as those explained in the subsection \ref{subsec22}. The implementation of boundary conditions is the same as what is explained in the subsection \ref{subsec23} (a-coefficients correction). In what follows, we will explain the iterative solution procedure. An iterative Newton-like method should be used because of the nonlinearity that exists in Eq. \eqref{eq22}. We chose the basic Newton's method, which is known for its quadratic convergence, as 
\begin{equation}
\label{eq24}
\vec{\phi}^{(k+1)}=\vec{\phi}^{(k)}-[J(\vec{\phi}^{(k)})]^{-1} ~\vec{d}~(\vec{\phi}^{(k)}),
\end{equation}
where the superscript \textit{(k)} denotes the number of iteration and $[J(\vec{\phi})]$ is the Jacobian matrix.  Jacobian matrix is given by 
\begin{eqnarray}
\label{eq25}
\frac{\partial{\vec{d}(\vec{\phi})}}{\partial\vec{\phi}}=
[A]+\frac{\nu_v\nu_sqL_{sc}^2}{2\epsilon_0 V_T}diag
\Big( \vec{N_C}\vec{\mathcal{F}}_{-\frac{1}{2}}(\vec{\phi}) \Big).~~
\end{eqnarray}
where the $diag$ refers to a diagonal matrix.
The iterative process to solve Eq. \eqref{eq23} is depicted in Fig. \ref{fig3_flow}(a) in \ref{AppB}. 
The initial guess to start Newton's method can be taken as a zero vector, $\vec{\phi}^{~(k=0)}=\vec{0}$. We call the final electrostatic potential output $\vec{\phi}_{~TF}$.  

\subsection{Enforce self-consistent convergence}  
\label{subsec27}
In subsections \ref{subsec22} and \ref{subsec23}, we have elaborated on solving the Schr{\"o}dinger and Poisson's equations as independent equations. Subsequently, the self-consistency between the two main space-dependent parameters [$\phi(y,z)$ and $n(y,z)$] must be established. Since early works at the late 60s, several iteration procedures have been employed to satisfy the self-consistency such as under-relaxation and adaptive relaxation methods~\cite{kerkhoven1990efficient}, the perturbation method~\cite{tan1990self, stern1970iteration} and the predictor-corrector method~\cite{trellakis1997iteration}. The under-relaxation method is simple but this method is extremely time-consuming, and it is also prone to divergence when it comes to 2D and 3D geometries. We, therefore, have employed the \textit{predictor-corrector} to enforce convergence on the self-consistent loop. The implementation of this method divides into two steps, \textit{predictor} and \textit{corrector} steps.  

In the predictor step, the solution for a nonlinear Poisson-like (predictor-Poisson's) equation provides a better prediction of the correct electrostatic potential. The predictor step requires at least three input parameters (guesses): (I) electrostatic potential $V^{in}(y,z)$, (II) the corresponding eigenvalues $e_i$, and (III) the corresponding eigenvectors $\psi_i(y,z)$. In addition, one needs the profile of electron density, calculated from parameters (II) and (III), if the exchange-correlation potential is included in the problem. The scaled version of predictor-Poisson's equation is as follows:
\begin{eqnarray}
\begin{aligned}
\label{eq26}
&\bigl[
\frac{\partial }{\partial y_s} 
\Big(-\epsilon_{ry}(y,z)\frac{\partial }{\partial y_s}\Big)+ 
\frac{\partial }{\partial z_s} 
\Big(-\epsilon_{rz}(y,z)\frac{\partial }{\partial z_s}\Big)
\bigr] \\
&
\phi^{pr}(y,z)
=\frac{-\nu_v\nu_sqL^2_{sc}}{\epsilon_0}~\sum_{i}^{n_Q} n^{pr}_i,
\end{aligned}
\end{eqnarray}
where the superscript \textit{pr} is added for extra clarification and the integer $n_Q$ indicates the total number of included quantum states. Each source term, $n^{pr}_i(y,z)$, has a similar expression as the electron density in Eq. \eqref{eq6} as: 
\begin{eqnarray}
\begin{aligned}
\label{eq27}
n^{pr}_i &=\frac{\sqrt{m_x(y,z)}}{\sqrt{2\pi}~\hbar}
{(k_BT)}^{\frac{1}{2}} |\psi_i(y,z)|^2 \times \\
&
~~~~\mathcal{F}_{-\frac{1}{2}}\Big(\frac{\mu_e-e_i+(\phi^{pr}(y,z)-V^{in}(y,z))}{V_T} \Big),\\
\end{aligned}
\end{eqnarray}
except that, the term $(\phi^{pr}(y,z)-V^{in}(y,z))$ is added to the numerator of the Fermi-Dirac integral of the order -1/2. Parameters $V_T$, $\mu_e$, and $e_i$ are in the eV unit. 
We denote the numerical solution of Eq. \eqref{eq26} as $\vec{\phi}_{{PR}}^{~out}$
and the residue between the input and the output electrostatic potentials as ${d}^{sc}(y,z)={\phi}_{PR}^{out}(y,z)-{V}^{in}(y,z)$. The predictor-Poisson's equation is derived based on the first-order perturbation theory and it predicts a better estimation for the electrostatic potential as a function of the three aforementioned guesses on the previous cycle~\cite{trellakis1997iteration}. Therefore, it is expected that ${d}^{sc}$ reduces gradually. 

In the corrector step, a new set of eigenvalue and eigenfunction (with $n_Q$ member) is calculated by solving the scaled Schr{\"o}dinger equation Eq. \eqref{eq15}, in which $\phi(y,z)={\phi}_{PR}^{out}(y,z)$. 
Thus, eigenvalues and eigenfunctions are corrected in this step. Note that, the profile of exchange-correlation potential, $V_{xc}(n(y,z))$, is also needed if electron-electron interaction is included. The cycle of correction and prediction are interchangeable, and two steps repeat until the condition $max|{d}^{sc}(y,z)|<V_{SC}^{tol}$ is met. The $V_{SC}^{tol}$ refer to a satisfactory low value, for instance $10^{-6}~V$. 
So far, we explained the predictor-corrector method starting from the predictor step. However, It seems more convenient to start from the correction step because one needs only a guess for the profile of electrostatic potential [$\phi(y,z)=V^{in}(y,z)$], rather than guesses for sets of paired eigenvalues and eigenfunctions as well as the electrostatic potential needed in the predictor step.  

Importantly, it has been confirmed by several reports that further reduction in the number of self-consistent iterations is possible by employing a method called \textit{Anderson mixing}~\cite{eyert1996comparative,wang2009accelerated,gao2014efficient}. To summarize, we have employed a combination of the predictor-corrector method and Anderson mixing, and we have started from the correction step. With that, \textit{corrector} and \textit{predictor} steps can be called \textit{outer} and \textit{inner} steps (loops) in the self-consistent level, see loops in Fig. \ref{fig3_flow}(b) in \ref{AppB}. 

\subsection{Numerical implementation of Finite-Volume} 
\label{subsec28}
In order to attain the self-consistent solution, the Finite-Volume discretized form of the predictor-Poisson's equation should be derived. To do so, an estimation for the integration of the highly nonlinear source terms [on the right-hand side of Eq. \eqref{eq27}] over a CV is given as
\begin{eqnarray}
\begin{aligned}
\label{eq28}
&\int_{z_s}^{z_n}\int_{y_w}^{y_e} \sum_{i} n^{pr}_i=
\sum_{i}\int_{z_s}^{z_n}\int_{y_w}^{y_e} N_{C_{1D}}|\psi_i|^2  \times \\
&
\mathcal{F}_{-\frac{1}{2}}
\Big(\frac{\mu_e-e_i+(\phi^{pr}-V^{in})}{V_T} \Big) \approx 
{N_{C_{1D}}}_P \times \\
&
\sum_{i}|{\psi_i}_P|^2 \mathcal{F}_{-\frac{1}{2}}
\Big(\frac{\mu_e-e_i+({\phi^{pr}}_P-{V^{in}}_P)}{V_T} \Big) \Delta y \Delta z. 
\end{aligned}
\end{eqnarray}
In the above notation, $n^{pr}_i$, $N_{C_{1D}}$, $\psi_i$, $\phi^{pr}$ and $V^{in}$ (in the integrand) are space dependent. In the last line of Eq. \eqref{eq28}, these parameters are reduced to their average values associated with the cell's center, the $P$. The Finite-Volume form of predictor-Poisson's equation is similar to the form of Poisson's equation with the nonlinear Thomas-Fermi electron density as explained earlier, except here more parameters are involved in the nonlinear source term. Consequently, the residual form of the discretized predictor-Poisson's equation can be written in a matrix form as
\begin{eqnarray}
\begin{aligned}
\label{eq29}
& \vec{d}^{~pr~(k)}~( \vec{\phi}^{~pr~(k)} )=[A ] ~\vec{\phi}^{~pr~(k)}+\frac{\nu_v\nu_sq L^2_{sc}}{\epsilon_0}\vec{N}_{C_{1D}}\times
\\&
 \sum_{i}^{n_Q} |{\vec{\psi}_i^{~(O)}}|^2
\vec{\mathcal{F}}_{-\frac{1}{2}} 
\big( {e_i}^{(O)},\vec{\phi}^{~pr~(k)}, \vec{V}^{in~(O)} \big)
-\vec{R_B}=0.
\end{aligned}
\end{eqnarray}
The superscript \textit{(k)} denotes the number of iterations on the inner (prediction) loop, while the extra superscript \textit{(O)} represents the iteration number on the outer (correction) loop. To make the notation simple, we just kept, $\vec{\phi}^{~pr~(k)}$ (the dependent vector), $\vec{V}^{~in~(O)}$ (a fixed vector) and ${e_i}^{(O)}$ (a fixed value) in the function $\mathcal{F}_{-{1}/{2}}$. 
However, the Fermi-Dirac integral ($j=-1/2$) also depends on thermal energy $V_T$ and electrochemical potential $\mu_e$, see Eq. \eqref{eq28}. The matrix $[A]$ and the vector $\vec{R_B}$ are what we have seen in the last paragraph of the subsection \ref{subsec23} and in the process of implementation of boundary conditions on the subsection \ref{subsec24}. 
Eq. \eqref{eq29} should be solved with a Newton-like iterative method similar to the case that is explained earlier for Thomas-Fermi approximation. Thus, one needs the Jacobian matrix as:
\begin{eqnarray}
\begin{aligned}
\label{eq30}
&[J( \vec{\phi}^{~pr~(k)})]^{~k}=[A]- \frac{\nu_v\nu_sqL_{sc}^2}{2\epsilon_0 V_T} \times diag \\
& 
\Big(\vec{N}_{C_{1D}}\sum_{i} |{\vec{\psi}_i^{~(O)}}|^2
\vec{\mathcal{F}}_{-\frac{3}{2}} \big( {e_i}^{(O)},\vec{\phi}^{~pr~(k)},\vec{V}^{~in~(O)} \big)\Big).
\end{aligned}
\end{eqnarray}
The details of the workflow are shown in Fig. \ref{fig3_flow}(b). 

 \subsection{Self-validation}
\label{subsec29}
In general, it is possible to validate the correctness of the dependent variable when a Finite-Volume method is used to solve a PDE by applying Gauss's law. In the following, we will briefly explain the validation of $\phi(y,z)$. Gauss’s law for Poisson's equation is given by
\begin{equation}
\label{eq31}
\iint_{\Omega} \vec{\nabla} \cdot{\bf{D}}~  d{\bf{A}}= \oint_{\partial\Omega} {\bf{D}}\cdot {\bf{n}}ds =\frac{q}{\epsilon_0}\iint_{\Omega} n~d{\bf{A}},
\end{equation}
where ${\Omega}$, ${\partial\Omega}$, and ${\bf{n}}$ represent the domain's total area, the circumference of the domain, and the unit vector normal to the surface, respectively. The above relation can be written in the discretized form as
\begin{eqnarray}
\begin{aligned}
\label{eq33}
&\sum_{\{sB\}}D_{S}\Delta y+ 
\sum_{\{wB\}}D_{W}\Delta z+
\sum_{\{nB\}}D_{N}\Delta y+ \\
&
\sum_{\{eB\}}D_{E}\Delta z
=\frac{q}{\epsilon_0} \sum_P n_P \Delta y \Delta z
=\frac{q}{\epsilon_0} n_{1D},
\end{aligned}
\end{eqnarray}
where ${\{sB\}}$ (${\{wB\}}$, ${\{nB\}}$, ${\{eB\}}$) refers to a set of CVs which have Southern (Western, Northern, Eastern) boundary interfaces. In fact, the parameter $n_{1D}=\sum_P n_P \Delta y \Delta z$ is the one-dimensional electron density along the x-axis (with the unit of length$^{-1}$).  If the two sides of Eq. \eqref{eq33} are equal it means the solution process is correct. We simply call the difference between the left and right sides of Eq. \eqref{eq33} as: \textit{imbalance}. The lower the imbalance is the more valid the numerical process would be. 
 \section{Results and Discussions}
\label{sec3}
We have described the geometrical and material aspects of our MOS nanowire example at the beginning of this section. We also explained our strategy for achieving self-consistent convergence. Then, adopting the FV-SP as our primary simulation tool and FV-TF as the prerequisite for FV-SP, we explored the characteristics of the gate-defined Si nanowire at different temperatures. We also tracked the formation of subbands by sweeping the top gate voltage. The validity of FV-SP and FV-TF are explored at this stage via the feature of self-validation. The effect of electron-electron iteration on different characteristics of 1DEG is also investigated. 
\subsection{Device and solution convergence}
\label{subsec31}
The width of the top metal gate is considered to be $40~nm$, and it is biased with $V_g=1.00~V$ initially. The top gate is located symmetrically between two uncontacted $50~nm$ spaces along the y-axis on the top surface [see Fig. \ref{fig1}(b)]. Table (I) lists the material properties and thicknesses of the layers. 
\begin{table} [h]
\small
\centering
  \caption{\ Set of the material parameters.}
  \label{tbl:example1}
  \begin{tabular*}{0.49\textwidth}{@{\extracolsep{\fill}}llllll}
    \hline
   Layer &r$^{\star}_{x}$&r$^{\star}_{y}$&r$^{\star}_{z}$&
    $\epsilon_{x}=\epsilon_{y}=\epsilon_{z}$&$\chi_e(eV)$\\
    \hline
    Al$_2$O$_3$(5~nm) &0.40 &0.40 &0.40 &9.3   &3.4\\
    SiO$_2$(5~nm)     &0.58 &0.58 &0.58 &3.9   &3.7\\
    Si(40~nm)         &0.19 &0.19 &0.90 &11.6  &0.0\\
    \hline
  \end{tabular*}
\end{table}

The electrochemical potential reference was considered as $\mu=0.0~meV$ (grounded). $80~K$, $4.2~K$, and $50~mK$ have been used as references for high, low, and ultra-low temperatures. 

\begin{figure}[h!tb]
	\begin{center}
		\includegraphics[width=8cm]{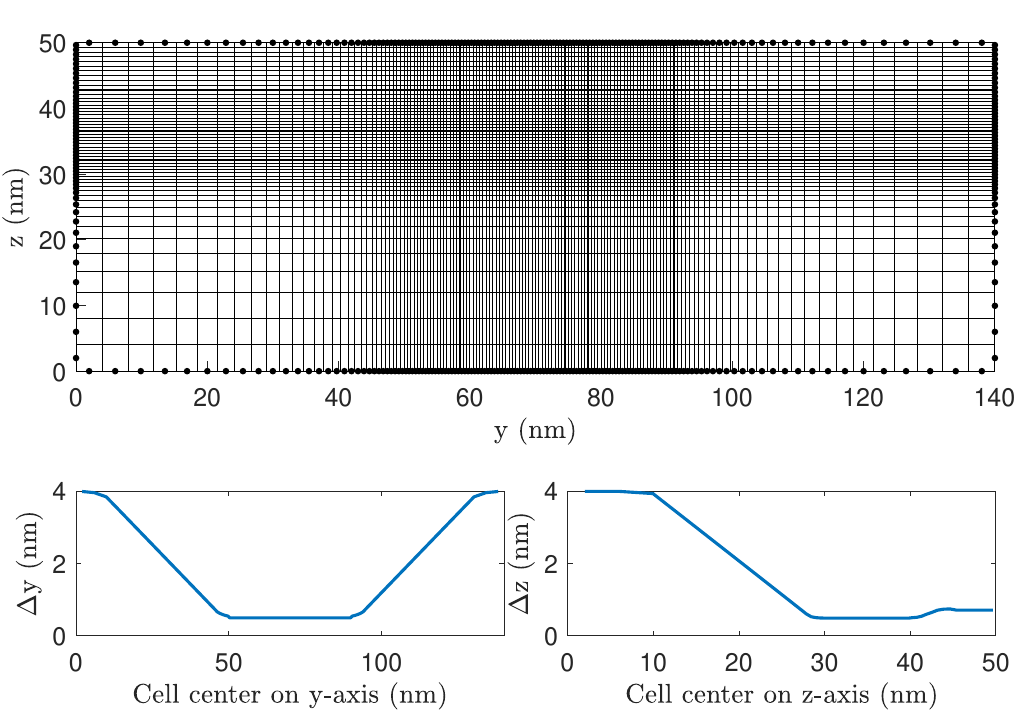}
	\end{center}
	\caption{\label{fig4} Upper panel shows the mesh used in our calculation. Small dots around the circumference are boundary points. Lower panels show how cell sizes change along the two axes. 
	}
\end{figure}

We developed our own Finite-Volume Thomas-Fermi approximation (FV-TF) and Finite-Volume Schr{\"o}dinger-Poisson (FV-SP) codes. Fig. \ref{fig4} shows the primary nonuniform mesh used in our calculations. In the real geometry, mesh sizes vary smoothly between a maximum of $4~nm$ and a minimum of $0.75~nm$ in both directions, such that mesh size becomes finer beneath the top gate, as shown in lower panels of Fig. \ref{fig4}. Note that in the scaled geometry, mesh sizes vary between $4~m$ and $0.75~m$. 
\begin{figure*}[]
	\begin{center}
		\includegraphics[width=16.0cm]{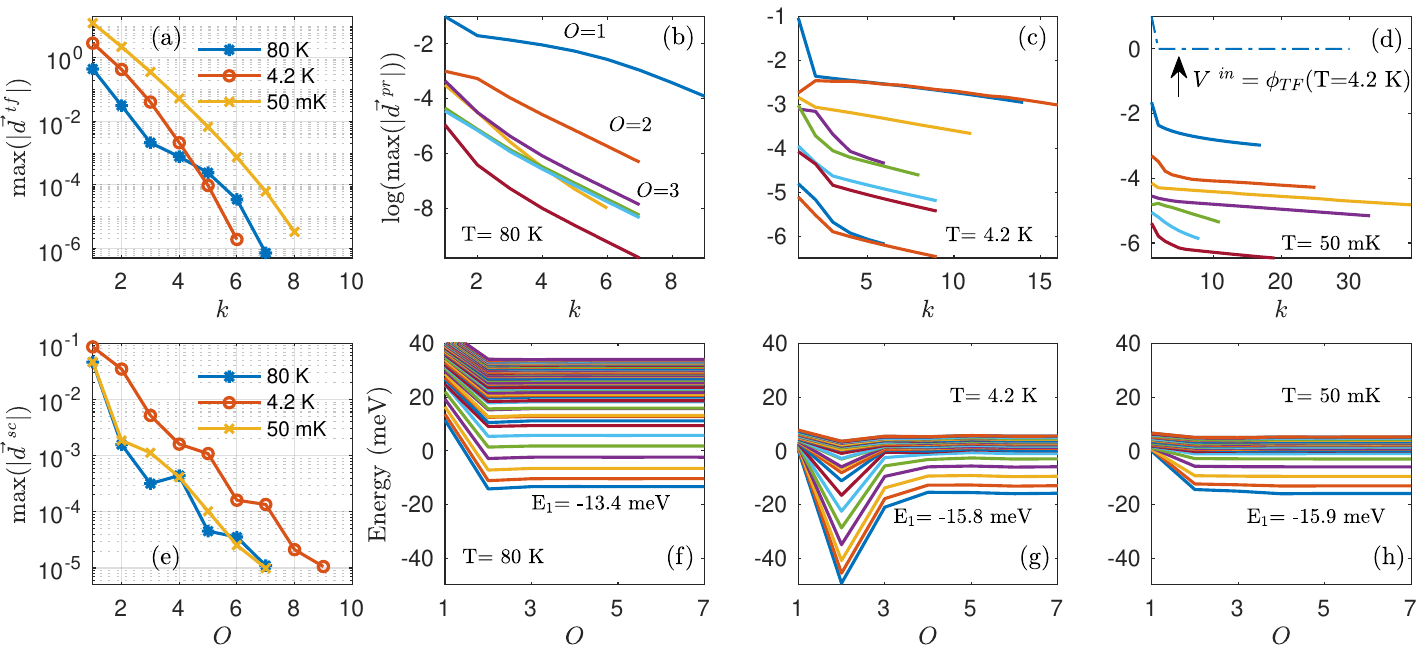}
	\end{center}
	\caption{\label{fig5} Convergence slopes for the electrostatic potential (a) calculated by FV-TF method, (b)-(d) calculated by inner loops of FV-SP method at three different temperatures. First slopes in (b)-(d) initiated with the corresponding outputs of the panel (a). In (b), the first few convergence slopes are labeled with the number of outer iteration $O$. The dot-dashed line in (d) is the convergence slope of the first inner loop when the $V^{in}$ is substituted by the output of the FV-TF method at a much higher temperature $T=4.2~K$. (e) Outer convergence slopes of the FV-SP method at three different temperatures. (f)-(h) Evolution of eigenvalues on the first seven outer cycles, associated with correction steps of (b)-(d).} 
\end{figure*}
\begin{figure*}[]
	\begin{center}
		 \includegraphics[width=16.0cm]{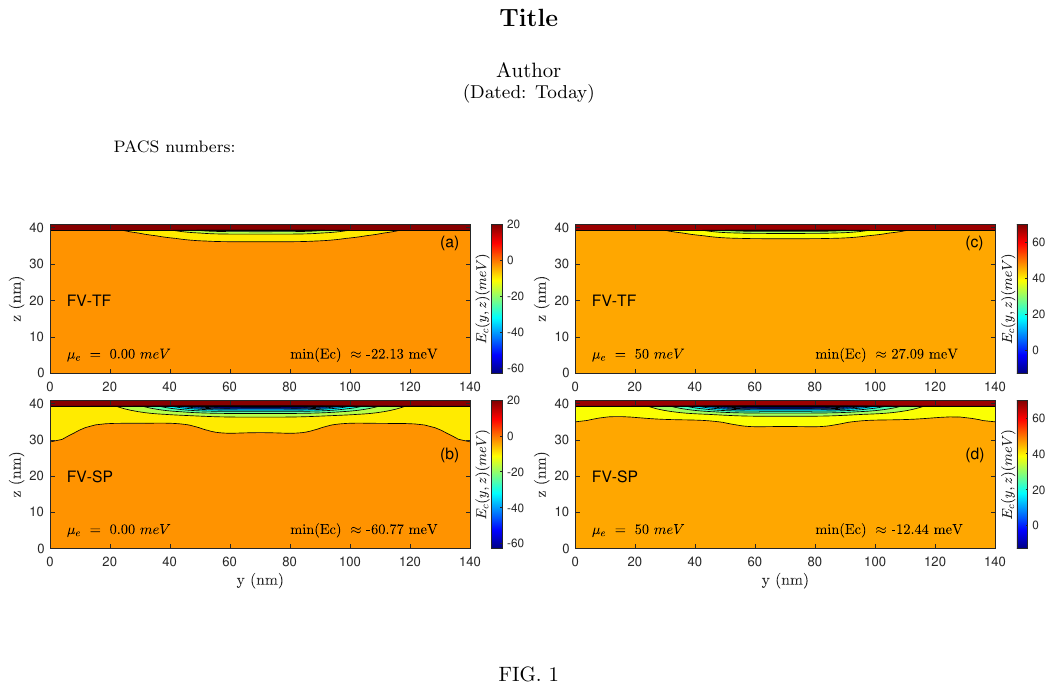}%
	\end{center}
	\caption{ Contour plot of the conduction band edge calculate (a) by FV-TF and (b) by FV-SP employing $V_g=1~V$ and $\mu_e=0.0~meV$ at $T=50~mK$. (c) and (d) Similar plots as (a) and (b) using a different electrochemical potential $\mu_e=50~meV$.} 
 \label{fig6}
\end{figure*}
Moreover, we set $\nu_s=2$ (spin) and $\nu_v=2$ which means only the lowest two-fold degenerate band $\Delta_2$ is considered (the crystal orientation [001] is along the z-axis)~\cite{baykan2010strain}.
Our FV-TF model exhibits excellent convergence behavior, from higher to ultra-low temperatures, as shown in Fig 5(a). Depending on the temperature, the FV-TF model converged very quickly in 6 to 8 cycles (within a few seconds). We set the breaking condition as $V^{tol}_{TF}=10^{-5}~V$.

Outputs, $\phi_{TF}$, and $n_{TF}$ (without electron-electron interaction) are employed as inputs in the FV-SP model. 
The FV-SP method was found sensitive to the total number of quantum states ($n_Q$), such that the FV-SP did not converge for $n_Q<40$ at $T=80~K$. The FV-SP method requires two loop-breaking conditions to be adjusted at the outer and inner loops. Initially, the $V_{SC}^{tol}=V_{PR}^{tol}=10^{-5}~V$ is chosen as the breaking conditions. At different temperatures, we have carefully examined the convergence behavior of the inner and outer loops of FV-SP for multiple top gate voltages and electrochemical potentials. It has been understood that it is not a good idea to break the inner loop with a constant condition, particularly at ultra-low temperatures. Hence, one important question is: what is the best breaking condition for inner loops and how should we set the $V_{PR}^{tol}$? 

Two reasons for not using a fixed breaking condition in the inner loop are as follows: (I) At the sub-Kelvin range, the numerical cost of predictor-Poisson's equation (inner loop) is high. One reason for the high cost of these calculations is the difficulty in numerically estimating the Fermi-Dirac integrals, of order $-1/2$ and $-3/2$, with high accuracy~\cite{mohankumar1995accurate,fukushima2015precise}. In practice, the inner loop should modify the input potential, $V^{in}$, only to a sufficient level, and then it is crucial to correct the eigenvalues and eigenfunctions. (II) We cannot precisely predict the correctness of eigenvalues and eigenfunctions in each self-consistent loop. Thus, a fixed breaking condition in the inner loop imposes unnecessary costs on the calculations. In order to improve this, we have employed a dynamical method to break the inner loop. In our new method, each inner loop is repeated at least six times and the residual vectors, $\vec{d}^{~pr~(i)}$, are stored. The inner loop breaks when the condition $V_{PR}^{tol}=\mathcal{L}\times max|\{\vec{d}^{~pr~(3)},...,\vec{d}^{~pr~(6)}\}|$ is met. The $\mathcal{L}$ is a fixed value, which we refer to as \textit{lowering}. The $max|\{\vec{d}^{~pr~(1)}~,\vec{d}^{~pr~(2)}\}|$ is ignored, since the slope of convergence for the first two inner iterations is not monotonic. We discovered that a monotonic convergence slope is correlated with more accurate predictions for the electrostatic potential. Multiple inner convergence slopes are plotted in Figs. \ref{fig5}(b)-(d) at different temperatures. The first few convergence slopes in Fig. \ref{fig5}(b) are labeled with the number of the outer loop, $O$. 
We found that a number in the range between $10^{-2}$ [Fig. \ref{fig5}(b)] and $10^{-0.5}$ [Figs. \ref{fig5}(c) and \ref{fig5}(d)] provides an optimum value for the $\mathcal{L}$. At the few Kelvins range, employing a $\mathcal{L}$ less than $10^{-0.5}$ increases the number of outer iterations, reflecting insufficient modification on outputs of the inner loop. 

Careful examinations reveal that the good trend on the convergence of inner loops is dominant only when the same temperature is taken into account in the calculation of the initial guess ($V^{in}=\phi_{TF}$) by FV-TF model. In particular, if $\phi_{TF}$ of a higher temperature utilizes as $V^{in}$ in  the inner loop, then the inner loop's convergence slope is not quadratic, as shown by a dot-dashed line in Fig. \ref{fig5}(d). In this case, the slope of convergence is so small, such that the first inner convergence slope ($O=1$) looks flat and it is less likely to be converged with an acceptable number of iterations. This problem arises from the fact that the convergence of Newton's method depends heavily on the initial guess for the highly nonlinear PDEs~\cite{aggarwal2020beyond}. This shows the importance of providing a better initial guess by solving the FV-TF at much lower temperatures. Self-consistent convergence slopes for different temperatures are plotted in Fig. \ref{fig5}(e).
For considered temperatures, the evolution of the few lowest eigenvalues along the correction step is plotted in Figs. \ref{fig5}(f)-(h). It can be understood that substantial corrections on eigenvalues happen only after the first or second cycle of the self-consistent loop. In addition, the breaking condition $V_{SC}^{tol}=10^{-5}~V$ provides sufficient accuracy to get a stable set of subband energy. The subband energies noticeably move downward as the temperature is reduced from $T=80~K$ to $T=4.2~K$. However, the subband energies do not move any further as the temperature is lowered from $T=4.2~K$ to $T=50~mK$. 

\subsection{Characteristics of 1DEG and solution verification}
\label{subsec32}
We first focus on comparing the conduction band edges, $Ec(y,z)$, calculated by FV-TF and FV-SP methods at $50~mK$. Color contour plots of $Ec(y,z)$ are shown in Figs. \ref{fig6}(a) and \ref{fig6}(b). 
The lowest value of the conduction band edge is denoted by $min(Ec)$ (the bottom of the well) within each panel. The $Ec(y,z)$ calculated by FV-SP shows much softer spatial variations around the quantum well area. The area (shape) of the quantum well can be determined roughly by those (x,y)s where $Ec(x,y)<\mu_e$. The shape of the quantum well corresponds to the physical distribution of 1DEG. The longitudinal distribution of the quantum well is much larger (about $20~nm$ from each side) than the top gate width (40~nm). These results are strongly against using the hard-wall approximation on the y-axis. Readjusting the electrochemical potential to $\mu_e=50~meV$, the $Ec(y,z)$ is calculated again and the results are plotted in Figs. \ref{fig6}(c) and \ref{fig6}(d). Convergence behavior showed very little sensitivity to the value of the electrochemical potential. Comparison between Fig. \ref{fig6}(b) and Fig. \ref{fig6}(d) [Fig. \ref{fig6}(a) and Fig. \ref{fig6}(c)] indicates that the conduction band on the Si layer ($z<40~nm$) moves upward almost identical to the $\mu_e$. The $Ec(x,y)$ calculated by both FV-SP and FV-TF methods show a similar linear response to the $\mu_e$. 

\begin{figure}[h!tb]
	\begin{center}
		\includegraphics[width=8.2cm]{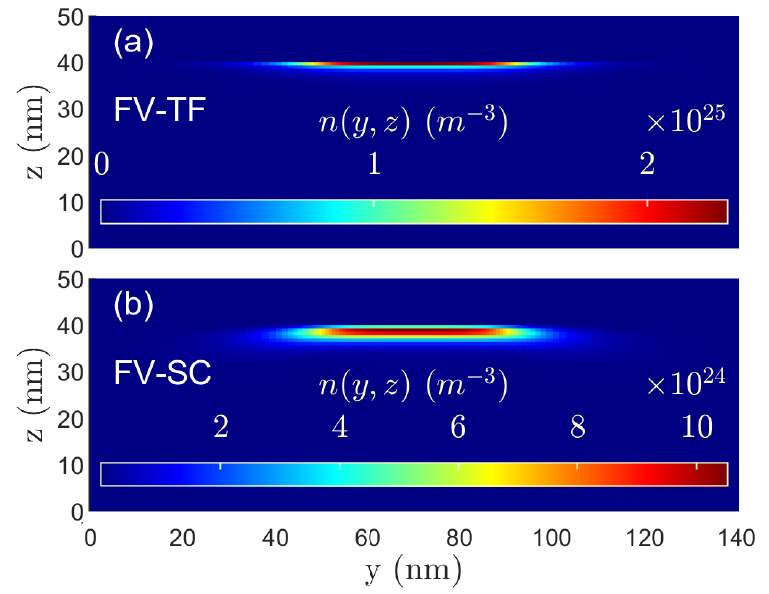} 
	\end{center}
	\caption{\label{fig7} (a) and (b) Show the electron densities calculated by FV-TF and FV-SP methods employing $V_g=1~V$ and $\mu_e=0.0~meV$ at $T=50~mK$.} 
\end{figure}

After characterizing the $Ec(x,y)$, we next explore the electron density. The $n(x,y)$ calculated by FV-TF and  FV-SP methods are plotted in Figs. \ref{fig7}(a) and \ref{fig7}(b). Note that, $n(x,y)$ is expressed with the unit $m^{-3}$ since we factored out the nanometer scale in the scaling process (see subsection \ref{subsec24}). Unpleasantly, $n(x,y)$ calculated by the FV-TF method is localized within a very narrow spatial interval along the z-axis with sharp convex edges along the y-axis. Whereas, the $n(x,y)$ calculated by the FV-CS method is localized along the z-axis nicely with smooth concave edges along the y-axis. The maximum of $n(x,y)$ calculated by the FV-TF method is almost two times larger than the same quantity that is calculated by the FV-SP method. 

\begin{figure}[]
	\centering
	\includegraphics[width=8.25cm]{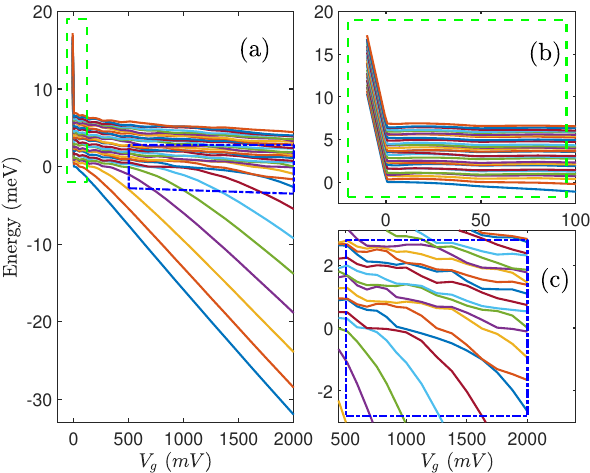}
	\caption{\label{fig8} (a) Formation of bound states (i.e., $e_i<\mu$) due to increase of the top gate voltage at $T=50~mK$ when $V_{xc}$ is excluded. (b) A magnified area of the (a) around $V_g=0~V$ where dense unbounded states are prominent. (c) A magnified area of the (a) where the excited states (i.e., $e_i>\mu$) responded nonlinearly to the top gate voltage.}
\end{figure}

Following the characterization of $Ec(x,y)$ and $n(x,y)$ at a fixed top voltage, we fix the temperature at $50~mK$ and sweep the top gate voltage over the range of $-10~mV<V_g< 2000~mV$, to examine the dependability of the FV-SP technique. This also allows us to study how changing the top gate voltage modifies the energy ladder. Figs. \ref{fig8}(a) and \ref{fig8}(b) show how bound states (i.e., states below the electrochemical potential, $e_i<\mu$) are formed from the dense eigenvalues of a particle in a large box (i.e., a particle in the $40~nm\times140~nm$ Si subdomain). As the top gate voltage rises, the lowest bound energy drops linearly. Upper-bound energies do not respond linearly to the top gate voltage. The unfilled states (i.e., excited states, the states above the Fermi-level $\mu=0~eV$) show anti-crossing features as shown in Fig. \ref{fig8}(c). 
\begin{figure}[]
	\centering
	\includegraphics[width=8.25cm]{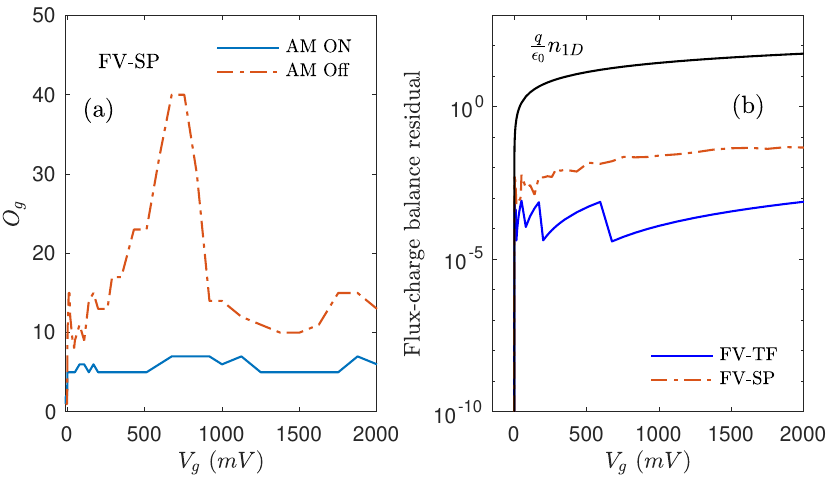}
	\caption{\label{fig9} (a) Number of outer iterations satisfying $V^{tol}_{SC}<10^{-5}~V$, vs the top gate voltage with and without the Anderson mixing (AM ON, AM Off) corresponding to the calculation made in Fig. \ref{fig8}. (b) Imbalances for FV-TF and FV-SP methods vs the top gate voltage. The $({q}/{\epsilon_0})n_{1D}$ represents the total flux out of the surface.}
\end{figure}

In addition, the number of necessary outer iterations, denoted by $O_g$, in the absence and presence of Anderson mixing are plotted in Fig. \ref{fig9}(a). For these calculations, the $\mathcal{L}=10^{-2.0}$ is selected to ensure the convergence of the inner loop for the entire spectrum of $V_g$.  In the absence (presence) of Anderson mixing, the $Q_{g}-V_g$ curve is achieved within about 40 hours (3 hours) on a four cores, Intel core i7 CPU with 16GB RAM. 
We stress two observations here: a) in the absence of Anderson mixing, the number of outer loop iterations shows a significant increase as the $V_g$ increases, as shown by the dot-dash line in Fig. \ref{fig9}(a). b) the Anderson mixing provide excellent assistance on reduction of the number of outer iterations, as shown by the solid line in Fig. \ref{fig9}(a). 
As a post-processing step, the imbalance, which represent the self-validation, is plotted in Fig. \ref{fig9}(b). In this stage, we also plotted $({q}/{\epsilon_0})n_{1D}$ in Fig. \ref{fig9}(b) to provide a quantitative reference for the total flux out of surfaces [see Eq. \eqref{eq33}]. One can see that  the imbalance, is approximately three orders of magnitude smaller than the total flux out of the surface. The FV-TF approach, with only one nonlinear equation, provides a better balance (lower imbalance) whereas FV-SP, with two PDEs, has a larger imbalance. The higher imbalance for the FV-SP method is rational because an extra imbalance originates from the numerical solution of the Schr{\"o}dinger equation. In practice, two factors can reduce the imbalance: one is a finer mesh and the other one is a lower value for $V^{tol}_{SC}$.

\begin{figure}[]
	\centering
	\includegraphics[width=8.25cm]{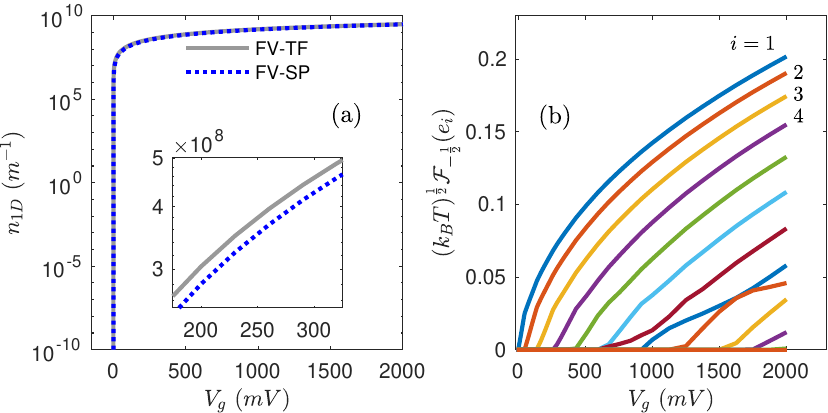}
	\caption{\label{fig10} (a) One-dimensional electron density, $n_{1D}$, vs the top gate voltage, $V_g$, for both FV-TF and FV-SP. Inset shows the Thomas-Fermi approximation overestimates the $n_{1D}$. (b) Evolution of the factor $(k_BT)^{1/2}\mathcal{F}_{-1/2}(e_i)$ under the influence of $V_g$. In (b), the first four lowest states are labeled by integer numbers.}
\end{figure}

Channel's switching property can be understood by plotting the one-dimensional electron density, $n_{1D}$, as a function of external gate voltage, $V_g$, as shown in Fig. \ref{fig10}(a). The ideal 1DEG quickly turns ON due to very thin layers of oxides (in total $10~nm$). Note that, we exclude extra sources of charge (such as doping on the silicon layer or dipoles on interfaces). An important point can be understood from the inset in Fig. \ref{fig10}(a), which is Thomas-Fermi approximation overestimates the $n_{1D}$ in comparison with the self-consistent approach. This is expected since the FV-TF method delivers a larger electron density as compared to the FV-SP method [see the maximum values in Figs. \ref{fig7}(a) and \ref{fig7}(b)]. Our calculated $n_{1D}$ is consistent with the reported 1D electron density which was obtained in the simulation of cylindrical silicon nanowire transistors~\cite{wang2004three}. In addition, our calculated switching property for the electron density is consistent with the transfer characteristics of n-MOS measured at deep-cryogenic temperatures~\cite{beckers2018characterization}. Furthermore, the evolution of the factor $(k_BT)^{1/2}\mathcal{F}_{-1/2}(e_i)$ due to the increase of the top gate voltage are plotted in Fig. \ref{fig10}(b).

\subsection{Effect of electron-electron interaction}
\label{subsec33}
\begin{figure}[]
	\centering
	\includegraphics[width=8cm]{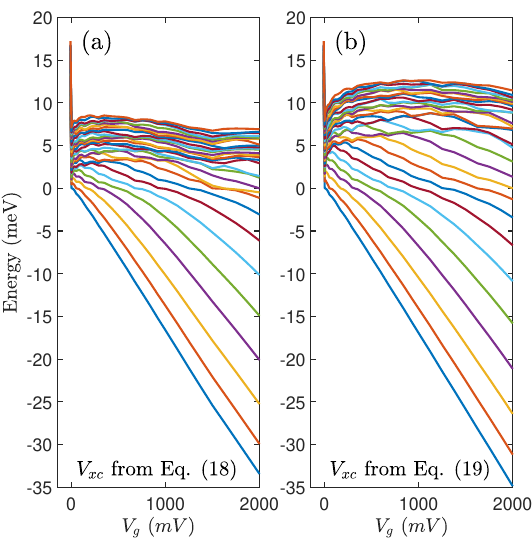}
	\caption{\label{fig11} (a) Formation of bound and excited states in the presence of exchange-correlation functional of Eq. \eqref{eq18} at $50~mK$. (b) Same as (a) but considering the exchange-correlation functional of Eq. \eqref{eq19}.}
\end{figure}

In the last part of the results, we incorporate two types of exchange-correlation functionals, mentioned in Eqs. \eqref{eq18} and \eqref{eq19}, into the FV-SP method and sweep the top gate voltage at $50~mK$. Eq. \eqref{eq18} is known as  the Hedin-Lundqvist functional. The Anderson mixing maintains its good performance in the presence of both exchange-correlation functionals. It is worth reporting that the Hedin-Lundqvist functional produces an instability spike only on the convergence slope of the first inner loop at ultra-low temperatures (a spike on $O=1$ in the Fig. \ref{fig5}(b), not shown here). This spike on the convergence slope is  suppressed toward the end of the first inner loop. We did not observe any serious divergence behavior such as fluctuation in the convergence rate of the inner loops at various top gate voltages. Formation of bound and excited states (i.e., states above the electrochemical potential $e_i>\mu$) considering Eq. \eqref{eq18} and Eq. \eqref{eq19} are plotted in Figs. \ref{fig11}(a) and \ref{fig11}(b), respectively. 
The Hedin-Lundqvist functional [Eq. \eqref{eq19}] affects the $e_i-V_g$ relation more than the functional of Eq. \eqref{eq18}. Both exchange-correlation functionals reduced the lowest bound energies by a few $meV$ [compare Figs. \ref{fig11}(a) and \ref{fig11}(b) with Fig. \ref{fig6}(a)]. Adding an exchange-correlation function also induces energy separation between excited states. In addition, the exchange-correlation of Eq. \eqref{eq19} produces more pronounced anti-crossing features for $e_i>\mu$. 

\begin{figure}[]
	\centering
	\includegraphics[width=8cm]{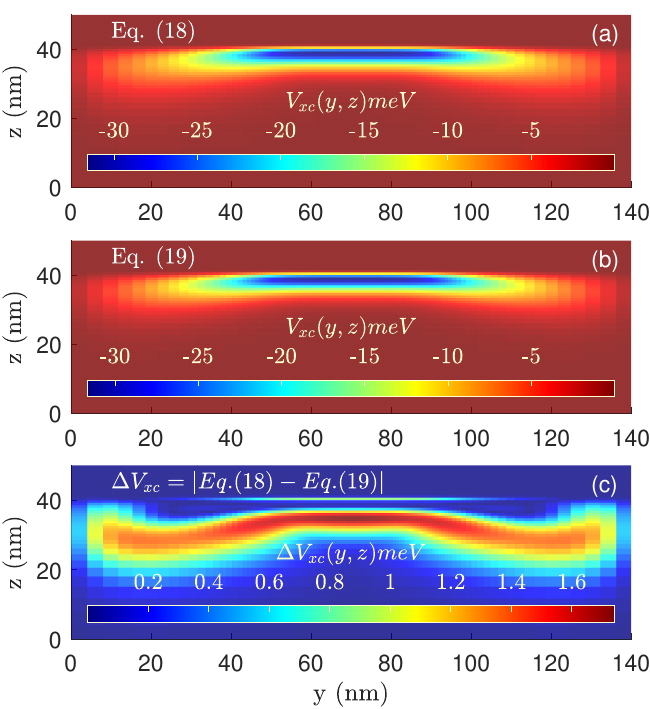}
	\caption{\label{fig12}(a) Spatial variation of exchange-correlation functional evaluated by Eq. \eqref{eq18} at $V_g=1~V$ and $T=50~mK$. Same as (a) evaluated by Eq. \eqref{eq19}. (c) Difference between two relations of exchange-correlation functional. }
\end{figure}

To further quantify the difference between exchange-correlation functionals $V_{xc}$, we plotted the two exchange-correlation functionals in Figs. \ref{fig12}(a) and \ref{fig12}(b), (for $V_g=1~V$ at $T=50~mK$). The lowest values of $V_{xc}$ are almost identical, $min(V_{xc})=-30~meV$. The difference between these two functionals is plotted in Fig. \ref{fig12}(c). There is an insignificant difference between these two types of exchange-correlation potentials. 
\section{Conclusion}
\label{sec4}
In summary, we have proposed a combination of cell centered Finite-Volume Thomas-Fermi (FV-TF) and Finite-Volume Schr{\"o}dinger-Poisson (FV-SP) methods as an approach for modeling low-dimensional semiconductor devices. For a MOS heterostructure, the theory and implementation are explained. This approach has several advantages. The most significant benefit of the FV-SP method is that it offers robust numerical stability enabling us to do the calculation at sub-kelvin temperatures. Our approach employs a structured nonuniform mesh, which makes the technique suitable for mesoscopic systems. The convergence properties of the FV-SP method and the influence of initial conditions on its convergence are fully explored. Importantly, the implementation of flux continuities at local levels enables us to validate the process of calculation. 
With the MOS example, we have characterized different parameters of the 1DEG formed by biasing a top gate. The FV-SP method, featured by Anderson mixing, exhibits consistent convergence characteristics with respect to external gate voltage and electrochemical potential. Additionally, it has been shown that the electron distribution has relatively long tails beneath the oxide layers, which is against using the hard wall approximation in the direction perpendicular to the heterojunction formation. The FV-SP method allows us to incorporate exchange-correlation functionals. It has been shown that two frequently used exchange-correlation functionals have negligible differences from each other. Based on our analysis, the exchange-correlation function affects the excited states more than the bound states. 
Under extremely low temperatures, the presence of an exchange-correlation function has a minor effect on the one-dimensional electron density, $n_{1D}$. An interesting piece of evidence is that the $n_{1D}$ calculated by the FV-TF method is surprisingly close to that calculated by the FV-SP method. 
The current work lacks a direct comparison between FV-SP with Finite-Difference SP (FD-SP) and Finite-Element SP (FE-SP). In the future, we aim to compare the convergence performance of FV-SP with FD-SP and FE-SP for realistic structures.
\section*{Conflict of interest}
There are no conflicts to declare.

\section*{Data availability statement}
The data that support the findings of this study are available
upon reasonable request from the authors.

\section*{Acknowledgements}
\label{sec5}
We thank Wei Zhu and Kun Yang for very useful conversations. V.M acknowledges funding from Summer Academy Program for International Young Scientists (Grant No. GZWZ[2022]019). W.D acknowledges the startup funding from Westlake University.


\appendix
\section{Enforce the continuity of fluxes}
\label{AppA}
\begin{figure}[h]
		\begin{center}
			\includegraphics[width=6cm]{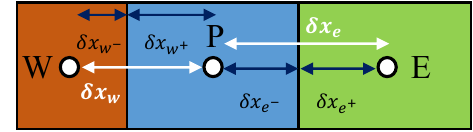}
		\end{center}
		\caption{\label{FS1} A schematic of three FV cells. Continuity of fluxes on eastern and western interfaces can be established by taking an averaging rule for material properties which requires the knowledge of relative distances between cell centers and interfaces.}
\end{figure}

It is very important to realize that we enforce the universal conservation law by integrating as the first step of Finite-Volume discretization. Local conservation laws need to be enforced by proper incorporation of material properties into a-coefficients as it is described in the following. There are two sets of data in the cell-centered Finite-Volume method: (I) cell-center data which refers to the average values associated with the cell centers, depicted by hollow circles in Fig. \ref{FS1}. (II) Nodal data refers to the data on the interfaces (or intersections). In the process of discretization, a precise connection between these two sets of data must be established based on the conservation laws. 

We take the electrostatic potential, $\phi$, as the dependent variable and the horizontal axis as the x-axis. We then focus on the eastern interface, the border between the central cell (the P) and the eastern cell (the E), see Fig. \ref{FS1}. The continuity of fluxes at the eastern interface of the cell P requires
\begin{equation}
\label{eqs1}
\Gamma_e\frac{\partial \phi}{\partial x}\biggr\rvert_{xe} = \Gamma_P\frac{\partial \phi}{\partial x}\biggr\rvert_{xe^-}=\Gamma_E\frac{\partial \phi}{\partial x}\biggr\rvert_{xe^+}.
\end{equation}
We remind that quantities that are defined by uppercase subscripts refer to the cell-center data whereas quantities that are defined by lowercase subscripts refer to the nodal data (on the interfaces between cells). The $\Gamma$ is the dielectric constant on Poisson's equation. On the other hand, if  we take the wavefunction, $\psi$, as the dependent variable then the $\Gamma$ is the diffusion coefficient of the Schr{\"o}dinger equation (which is proportional to the inverse of effective-mass). We emphasize that the continuity of the dielectric constant or the effective-mass are disregarded. Whereas, the qualities such as Eq. \eqref{eqs1} should be considered. The above relations are approximated by the central difference approximation as
\begin{eqnarray}
\begin{aligned}
\label{eqa2}
&\Gamma_e\frac{ \phi_E-\phi_P}{\delta x_e} =\Gamma_P\frac{ \phi_e-\phi_P}{\delta x_{e^-}}, \\
&
\Gamma_e\frac{\phi_E-\phi_P}{\delta x_e} =\Gamma_E\frac{\phi_E-\phi_e}{\delta x_{e^+}}.
\end{aligned}
\end{eqnarray}
The two above equality can be written as
\begin{eqnarray}
\begin{aligned}
\label{eqa3}
&\frac{\Gamma_e}{\delta x_e} 
\frac{\delta x_{e^-}}{\Gamma_P} \left(\phi_E-\phi_P\right) =\left(\phi_e-\phi_P\right),  \\
&
\frac{\Gamma_e}{\delta x_e} 
\frac{\delta x_{e^+}}{\Gamma_E} \left(\phi_E-\phi_P\right) =\left(\phi_E-\phi_e\right).
\end{aligned}
\end{eqnarray}

The $\phi_e$ will be canceled out by adding the two above equations as 
\begin{equation}
\label{eqa4}
\frac{\Gamma_e}{\delta x_e} 
\left( 
\frac{\delta x_{e^-}}{\Gamma_P} +
\frac{\delta x_{e^+}}{\Gamma_E} 
\right)
\left(\phi_E-\phi_P\right) = \left(\phi_E-\phi_P\right).
\end{equation}
Then, the $\Gamma_e$ reads as
\begin{equation}
\label{eqa5}
\Gamma_e
=\delta x_e
\left( 
\frac{\delta x_{e^-}}{\Gamma_P} +
\frac{\delta x_{e^+}}{\Gamma_E} 
\right)^{-1}.
\end{equation}
The $\Gamma_e$ can also be given by
\begin{equation}
\label{eqa6}
\Gamma_e =
\frac{\Gamma_E\Gamma_P}{\beta\Gamma_E+ (1-\beta)\Gamma_P},
\end{equation}
where $\beta=\delta x_{e^-}/\delta x_{e}$ and $(1-\beta)= \delta x_{e^+}/\delta x_{e}$. Eq. \eqref{eqa6} must be implemented during the evaluation of matrix coefficients to enforce the continuity of flux out of the surface. We call Eq. \eqref{eqa6} the \textit{opposite relative distance averaging}, since the $\beta$, a geometrical quantity belongs to the cell P, multiples to the $\Gamma_E$ which is a diffusion coefficient belongs to the opposite cell. The other multiplication $(1-\beta)\Gamma_P$ has a similar reverse fashion. Elsewhere, Eq. \eqref{eqa6} has been called \textit{Harmonic mean} and it guarantees the continuity of the flux out of surface at interfaces despite material discontinuities.
Subtracting the two equality in Eq. \eqref{eqa3} gives  
\begin{equation}
\label{eqa7}
\frac{\Gamma_e}{\delta x_e} 
\left( 
\frac{\delta x_{e^-}}{\Gamma_P} -
\frac{\delta x_{e^+}}{\Gamma_E} 
\right)
\left(\phi_E-\phi_P\right) =2\phi_e- \left(\phi_E+\phi_P\right).
\end{equation}
Modification of $\phi_e$ in terms of $\beta$, $\left(\phi_E-\phi_P\right)$
and $\left(\phi_E+\phi_P\right)$ and using Eq. \eqref{eqa6} give rise to
\begin{equation}
\label{eqa8}
\phi_e=
\left( 
\frac{\beta\Gamma_E-(1-\beta)\Gamma_P}
{\beta\Gamma_E+(1-\beta)\Gamma_P}
\right)
\frac{\phi_E-\phi_P}{2}+ 
\frac{\phi_E+\phi_P}{2}.
\end{equation}
We can use the translational method to derive the $\Gamma_w$ on the western interface as the following. We rename cells as P$\to$W and E$\to$P. This means the old Eastern interface is now the Western interface. Consequently, the same equation as Eq. \eqref{eqa6} can be used to enforce the continuity on the western interface as
\begin{equation}
\label{eqa9}
\Gamma_w = \frac{\Gamma_W\Gamma_P}{\beta'\Gamma_W+(1-\beta')\Gamma_P}.
\end{equation}
Here $\beta'= \delta x_{w^+}/\delta x_{w}$ and $(1-\beta')=\delta x_{w^-}/\delta x_{w}$. We have used the notation $\beta'$ to keep the consistency between Eq. \eqref{eqa6} and Eq. \eqref{eqa9} [i.e., the role of the opposite relative distance averaging]. Similarly, the $\phi_w$ can be calculated via the following relation
\begin{equation} 
\label{eqa10}
\phi_w=
\left( 
\frac{\beta'\Gamma_P-(1-\beta')\Gamma_W}
{\beta'\Gamma_P+(1-\beta')\Gamma_W}
\right)
\frac{\phi_P-\phi_W}{2}+ 
\frac{\phi_P+\phi_W}{2}.
\end{equation}
Eqs. \eqref{eqa8} and \eqref{eqa10} resemble to a weighted linear interpolation. Nodal data, $\phi_e$ and $\phi_w$ (or $\phi_n$ and $\phi_s$) do not play an essential role in our calculation. However, they can be useful in some cases. For instance, to approximate components of the directional electric field, $\partial\phi/ \partial x \approx (\phi_e-\phi_w)/\Delta x$  and $\partial\phi/ \partial y \approx (\phi_n-\phi_s)/\Delta_y$, as an average value for each cell. The same procedures in the y-axis give us similar expressions for $\Gamma_s$, $\Gamma_n$, $\phi_s$, and $\phi_n$.	

\section{Flowchart of the iterative process}
\label{AppB}
\begin{figure*}
	\begin{center}
		\includegraphics[width=14
		cm]{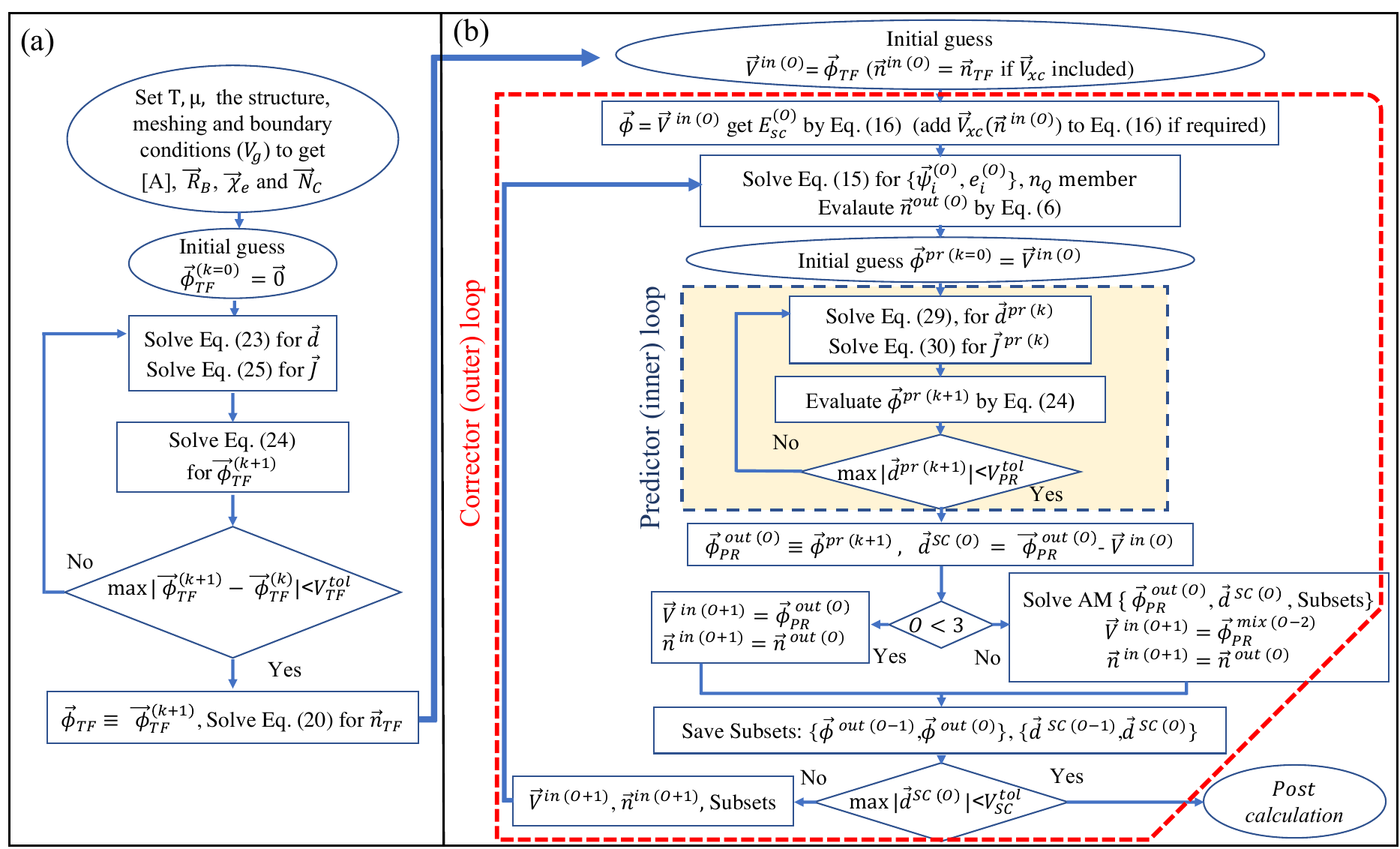}
	\end{center}
	\caption{\label{fig3_flow}  (a) Flowchart of the iterative process for Thomas-Fermi approximation. (b) Flowchart of our predictor-corrector method to solve Schr{\"o}dinger Poisson system.}
\end{figure*}

We have summarized the iterative processes for the Thomas-Fermi and our Finite-Volume self-consistent predictor-corrector methods in Fig. \ref{fig3_flow}. 

\newpage

\section*{References}
\bibliographystyle{iopart-num}
\bibliography{FVSC_BIB.bib}

\providecommand{\newblock}{}
\begin{thebibliography}{10}
\expandafter\ifx\csname url\endcsname\relax
  \def\url#1{{\tt #1}}\fi
\expandafter\ifx\csname urlprefix\endcsname\relax\def\urlprefix{URL }\fi
\providecommand{\eprint}[2][]{\url{#2}}

\bibitem{ando1982electronic}
Ando T, Fowler A~B and Stern F 1982 {\em Reviews of Modern Physics\/} {\bf 54}
  437

\bibitem{duke1967optical}
Duke C~{\'a} 1967 {\em Physical Review\/} {\bf 159} 632

\bibitem{bloss1989effects}
Bloss W~L 1989 {\em Journal of Applied Physics\/} {\bf 66} 3639--3642

\bibitem{datta2005quantum}
Datta S 2005 {\em Quantum transport: atom to transistor\/} (Cambridge
  university press)

\bibitem{yoshida1986classical}
Yoshida J 1986 {\em IEEE transactions on electron devices\/} {\bf 33} 154--156

\bibitem{woods2018effective}
Woods B~D, Stanescu T~D and Sarma S~D 2018 {\em Physical Review B\/} {\bf 98}
  035428

\bibitem{davies1998physics}
Davies J~H 1998 {\em The physics of low-dimensional semiconductors: an
  introduction\/} (Cambridge university press)

\bibitem{degtyarev2017features}
Degtyarev V, Khazanova S and Demarina N 2017 {\em Scientific reports\/} {\bf 7}
  1--9

\bibitem{wang2004three}
Wang J, Polizzi E and Lundstrom M 2004 {\em Journal of Applied Physics\/} {\bf
  96} 2192--2203

\bibitem{bell2010crossover}
Bell M, Sergeev A, Bird J, Mitin V and Verevkin A 2010 {\em Physical review
  letters\/} {\bf 104} 046805

\bibitem{stern1970iteration}
Stern F 1970 {\em Journal of Computational Physics\/} {\bf 6} 56--67

\bibitem{stern1972self}
Stern F 1972 {\em Physical Review B\/} {\bf 5} 4891

\bibitem{sarma1982electronic}
Sarma S~D and Vinter B 1982 {\em Physical Review B\/} {\bf 26} 960

\bibitem{lo1999modeling}
Lo S~H, Buchanan D~A and Taur Y 1999 {\em IBM Journal of Research and
  Development\/} {\bf 43} 327--337

\bibitem{duarte2010convergence}
Duarte C 2010 {\em Computer Physics Communications\/} {\bf 181} 1501--1509

\bibitem{tan1990self}
Tan I~H, Snider G, Chang L and Hu E 1990 {\em Journal of applied physics\/}
  {\bf 68} 4071--4076

\bibitem{ando2002numerically}
Ando T, Taniyama H, Ohtani N, Hosoda M and Nakayama M 2002 {\em IEEE journal of
  quantum electronics\/} {\bf 38} 1372--1383

\bibitem{nakamura1989finite}
Nakamura K, Shimizu A, Koshiba M and Hayata K 1989 {\em IEEE journal of quantum
  electronics\/} {\bf 25} 889--895

\bibitem{wu1993self}
Wu Z and Ruden P~P 1993 {\em Journal of applied physics\/} {\bf 74} 6234--6241

\bibitem{mazumder2015numerical}
Mazumder S 2015 {\em Numerical methods for partial differential equations:
  finite difference and finite volume methods\/} (Academic Press)
  (\textit{Preprint} \eprint{https://www.gettextbooks.com/isbn/9780128498941})

\bibitem{armagnat2019self}
Armagnat P, Lacerda-Santos A, Rossignol B, Groth C and Waintal X 2019 {\em
  SciPost Physics\/} {\bf 7} 031

\bibitem{berrada2020nano}
Berrada S, Carrillo-Nunez H, Lee J, Medina-Bailon C, Dutta T, Badami O,
  Adamu-Lema F, Thirunavukkarasu V, Georgiev V and Asenov A 2020 {\em Journal
  of Computational Electronics\/} {\bf 19} 1031--1046

\bibitem{baumgartner2013vsp}
Baumgartner O, Stanojevic Z, Schnass K, Karner M and Kosina H 2013 {\em Journal
  of Computational Electronics\/} {\bf 12} 701--721

\bibitem{angus2007gate}
Angus S~J, Ferguson A~J, Dzurak A~S and Clark R~G 2007 {\em Nano letters\/}
  {\bf 7} 2051--2055

\bibitem{brauns2018palladium}
Brauns M, Amitonov S~V, Spruijtenburg P~C and Zwanenburg F~A 2018 {\em
  Scientific reports\/} {\bf 8} 1--8

\bibitem{young1989position}
Young K 1989 {\em Physical Review B\/} {\bf 39} 13434

\bibitem{levy1995position}
Levy-Leblond J~M 1995 {\em Physical Review A\/} {\bf 52} 1845

\bibitem{burt1992justification}
Burt M 1992 {\em Journal of Physics: Condensed Matter\/} {\bf 4} 6651

\bibitem{bersch2008band}
Bersch E, Rangan S, Bartynski R~A, Garfunkel E and Vescovo E 2008 {\em Physical
  review B\/} {\bf 78} 085114

\bibitem{ribeiro2009accurate}
Ribeiro~Jr M, Fonseca L~R and Ferreira L~G 2009 {\em Physical Review B\/} {\bf
  79} 241312

\bibitem{stern1978image}
Stern F 1978 {\em Physical Review B\/} {\bf 17} 5009

\bibitem{snider1990electron}
Snider G, Tan I~H and Hu E 1990 {\em Journal of Applied Physics\/} {\bf 68}
  2849--2853

\bibitem{mizsei2002fermi}
Mizsei J 2002 {\em Vacuum\/} {\bf 67} 59--67

\bibitem{kamata2017design}
Kamata H and Kita K 2017 {\em Applied Physics Letters\/} {\bf 110} 102106

\bibitem{pacelli1997self}
Pacelli A 1997 {\em IEEE Transactions on electron devices\/} {\bf 44}
  1169--1171

\bibitem{heinz2004simulation}
Heinz F~O 2004 {\em Simulation approaches for nano-scale semiconductor
  devices\/} Ph.D. thesis ETH Zurich

\bibitem{ram2004schrodinger}
Ram-Mohan L, Yoo K and Moussa J 2004 {\em Journal of applied physics\/} {\bf
  95} 3081--3092

\bibitem{lether2000analytical}
Lether F~G 2000 {\em Journal of scientific computing\/} {\bf 15} 479--497

\bibitem{kim2008notes}
Kim R, Wang X and Lundstrom M 2008 {\em arXiv preprint arXiv:0811.0116\/}
  \urlprefix\url{https://arxiv.org/abs/0811.0116}

\bibitem{bednarczyk1978approximation}
Bednarczyk D and Bednarczyk J 1978 {\em Physics letters A\/} {\bf 64} 409--410

\bibitem{gao2013quantum}
Gao X, Nielsen E, Muller R~P, Young R~W, Salinger A~G, Bishop N~C, Lilly M~P
  and Carroll M~S 2013 {\em Journal of Applied Physics\/} {\bf 114} 164302

\bibitem{blazek2015computational}
Blazek J 2015 {\em Computational fluid dynamics: principles and applications\/}
  (Butterworth-Heinemann)

\bibitem{chen2010new}
Chen L 2010 {\em SIAM journal on numerical analysis\/} {\bf 47} 4021--4043

\bibitem{ando2003self}
Ando T, Taniyama H, Ohtani N, Nakayama M and Hosoda M 2003 {\em Journal of
  applied physics\/} {\bf 94} 4489--4501

\bibitem{hedin1971explicit}
Hedin L and Lundqvist B~I 1971 {\em Journal of Physics C: Solid state
  physics\/} {\bf 4} 2064

\bibitem{stern1984electron}
Stern F and Sarma S~D 1984 {\em Physical Review B\/} {\bf 30} 840

\bibitem{kerkhoven1990efficient}
Kerkhoven T, Galick A~T, Ravaioli U, Arends J~H and Saad Y 1990 {\em Journal of
  applied physics\/} {\bf 68} 3461--3469

\bibitem{trellakis1997iteration}
Trellakis A, Galick A, Pacelli A and Ravaioli U 1997 {\em Journal of Applied
  Physics\/} {\bf 81} 7880--7884

\bibitem{eyert1996comparative}
Eyert V 1996 {\em Journal of Computational Physics\/} {\bf 124} 271--285

\bibitem{wang2009accelerated}
Wang H, Wang G, Chang S and Huang Q 2009 {\em Micro \& Nano Letters\/} {\bf 4}
  122--127

\bibitem{gao2014efficient}
Gao X, Mamaluy D, Nielsen E, Young R~W, Shirkhorshidian A, Lilly M~P, Bishop
  N~C, Carroll M~S and Muller R~P 2014 {\em Journal of Applied Physics\/} {\bf
  115} 133707

\bibitem{baykan2010strain}
Baykan M~O, Thompson S~E and Nishida T 2010 {\em Journal of Applied Physics\/}
  {\bf 108} 093716

\bibitem{mohankumar1995accurate}
Mohankumar N and Natarajan A 1995 {\em physica status solidi (b)\/} {\bf 188}
  635--644

\bibitem{fukushima2015precise}
Fukushima T 2015 {\em Applied Mathematics and Computation\/} {\bf 259} 708--729

\bibitem{aggarwal2020beyond}
Aggarwal A and Pant S 2020 {\em Algorithms\/} {\bf 13} 78

\bibitem{beckers2018characterization}
Beckers A, Jazaeri F and Enz C 2018 {\em IEEE Journal of the Electron Devices
  Society\/} {\bf 6} 1007--1018

\end{thebibliography}

\end{document}